\documentclass[useAMS,usenatbib]{mn2e}

\usepackage{graphicx}

\newcommand{\xmm}{XMM-{\em Newton}}
\newcommand{\chandra}{{\em Chandra}}

\newcommand{\suzaku}{{\em Suzaku}}

\newcommand{\lognlogs}{Log~$N$--Log~$S$}

\newcommand{\de}{{\rm d}}
\newcommand{\ergs}{erg s$^{-1}$}

\newcommand{\e}[1]{\cdot 10^{#1}}

\newcommand{\chiq}{\chi^2}
\newcommand{\chiqr}{\chi^2_{\rm red}}
\newcommand{\ael}{$\alpha$-elements}

\newcommand{\lesssim}{\la}   
\newcommand{\gtrsim}{\ga}

\newcommand{\cSpvmkl}{{\tt c6pvmkl}}

\title[A deep X-ray observation of M82]{A deep X-ray observation of M82 with \xmm}
\author[P. Ranalli, A. Comastri, L. Origlia, R. Maiolino] 
       {P. Ranalli$^{1,2,3}$, A. Comastri$^{4}$, L. Origlia$^{4}$, R. Maiolino$^{5}$
         \thanks{E-mail: piero.ranalli@bo.astro.it (PR),
           an\-dre\-a.co\-ma\-stri@bo.astro.it (AC), 
           li\-via.o\-ri\-glia@bo.astro.it (LO),
           ma\-io\-li\-no@mporzio.astro.it (RM)}
         \footnotemark[1]
         \thanks{Based on observations obtained with XMM-Newton, an ESA science mission with instruments and contributions directly funded by ESA Member States and NASA.}\\
         $^{1}$RIKEN, Cosmic Radiation Laboratory,
         Hirosawa 2-1, Wakoshi, Saitama, 351-0198 Japan\\
         $^{2}$fellow of the Japan Society for the Promotion of Science (JSPS)\\
         $^{3}$(current address) Universit\`{a} di Bologna,
         Dipartimento di Astronomia, via
         Ranzani 1, 40127 Bologna, Italy\\
         $^{4}$INAF--Osservatorio Astronomico di Bologna, via Ranzani
         1, 40127 Bologna,
         Italy\\
         $^{5}$INAF--Osservatorio Astronomico di Roma, via di Frascati
         33, 00040 Monte Porzio Catone, Italy }
\begin{document}

\date{Accepted 2008/2/14. Received 2008/2/7; in original form 2007/9/14.}

\pagerange{\pageref{firstpage}--\pageref{lastpage}} \pubyear{2007}

\maketitle

\label{firstpage}

\begin{abstract}
  We report on the analysis of a deep (100 ks) observation of the
  starburst galaxy M82 with the EPIC and RGS instruments on board the
  X-ray telescope \xmm. The broad-band (0.5-10 keV) emission is due to
  at least three spectral components: i) continuum emission from point
  sources; ii) thermal plasma emission from hot gas; iii) charge
  exchange emission from neutral metals (Mg and Si). The
  plasma emission has a double-peaked differential emission measure,
  with the peaks at $\sim 0.5$ keV and $\sim 7$ keV. Spatially
  resolved spectroscopy has shown that the chemical absolute
  abundances are not uniformly distributed in the outflow, but are
  larger in the outskirts and smaller close to the galaxy centre. The
  abundance ratios also show spatial variations. The X-ray derived
  Oxygen abundance is lower than that measured in the atmospheres of
  red supergiant stars, leading to the hypothesis that a significant
  fraction of Oxygen ions have already cooled off and no longer emit
  at energies $\gtrsim 0.5$ keV.
\end{abstract}

\begin{keywords}
galaxies: individual: M82 -- galaxies: abundances -- X-rays: ISM --
ISM: jets and outflows -- plasmas -- atomic processes.
\end{keywords}

\section{Introduction}

M82 is a nearby galaxy often considered as the prototype starburst
galaxy \citep{rie80}. Its proximity \citep[3.63 Mpc,][]{distanza_m82} and
luminosity \citep[$10^{44}$ \ergs\ in the far infrared, $10^{40}$
\ergs\ in the X-ray domain; ][]{rcs03}, along with the presence of an
ongoing powerful ($\sim 3$ M$_\odot$ yr$^{-1}$) starburst has made it the
subject of countless studies at all wavelengths.  The starburst,
located in the galaxy central regions, is driving a several kpc large
outflow perpendicular to the plane of the galaxy \citep{heckman90}
which has been observed at multiple wavelengths, such as the radio
\citep{seaquist91}, infrared \citep{alton99,engelbracht07}, H$\alpha$
and X-rays \citep{lehnert99}.  Recent observations in the X-ray domain
with the \chandra, \xmm\ and \suzaku\ observatories are reported in
\citet{grif00}, \citet{rs02}, \citet{or04}, \citet{strickland07} and
\citet{tsuru07}. The small inclination of the galaxy disc makes M82
observed almost exactly edge-on, thus allowing a good perspective on
the supernova-driven outflow and contributing to its success as an
astrophysical laboratory.

M82 was thus chosen as a test case where the metal enrichment due to a
burst of star formation can be directly measured.  Determining the
abundance of key elements released in the interstellar medium (ISM) by
supernovae (SN) with different mass progenitors and hence on different
time scales, will have a strong astrophysical impact in drawing a
global picture of galaxy formation and evolution
\citep{mwi97,mc94}. While the metals locked into the stellar
atmospheres give a picture of the abundances at the beginning of the
last burst of star formation (SF), the hot gas heated by core-collapse
supernova explosions and emitting in the X-rays should trace the
enrichment by the SN which were formed in the new generation of
stars. 

In our previous paper \citep{or04} we presented high resolution
infrared (J and H band) spectra that we acquired with the 3.6 m
Italian Telescopio Nazionale Galileo (TNG), together with a rather
shallow ($\sim 15$ ks) X-ray observation obtained from the \xmm\
archive. Our results hinted for a confirmation of the expected
scenario in which the gaseous component has a higher content of
$\alpha$-elements than the stellar one, and a similar content of Fe.
However, some new issues were posed, since we found a very low
abundance of O and Ne with respect to other $\alpha$-elements (e.g.,
O/Mg $\sim 0.2$, Ne/Mg $\sim 0.3$) in the hot gas present in the
central ($\lesssim 1$ kpc) regions of M82, which could not be
satisfactorily explained. For these reasons we were granted the deeper
($\sim 100$ ks) observation of M82 which we present in this
article. Two reports on an early analysis appeared in
\citet{m82escorial06} and \citet{m82mssl06}. Both this work and
\citet{or04} are part of an ongoing larger effort to measure
metallicity enhancements in a sample of galaxies.

The X-ray spectrum of M82 is rich and complex.  For a review about the
physical and observational parameters which may be recovered through
X-ray spectroscopy (shape of the spectra; slopes of power-laws;
temperatures and abundances of thermal plasmas; etc.), we refer the
reader to \citet{paerels03}, while for the physical processes of
thermal and charge-exchange emission we refer to the book by
\citet{osterbrock-libro} and to the recent review by
\citet{kallman07}; for plasma emission we mention both historical
references such as \citet{landini70} and the current ones which
illustrate the state of the art (e.g., \citealt{mekal} and
\citealt{apec}).

This paper is structured as follows: Sect.~\ref{sec:observations}
presents the \xmm\ observation and data reduction of M82. In
Sect.~\ref{s:spettri1} the EPIC spectra of the central region are
analysed, and results are shown about the temperature distribution of
the plasma, charge exchange emission, and a comparison of different
plasma codes. In Sect.~\ref{sec:rgs} the RGS spectra are analysed,
improving and extending the results obtained from EPIC data. In
Sect.~\ref{sec:outflow} a spatially-resolved spectral analysis of the
EPIC data of the M82 outflow is presented; a dependence of the
chemical abundances on the height from the galaxy plane is found. The
implications of the main findings of this article are discussed in
Sect.~\ref{sec:discussione}.  Finally, in Sect.~\ref{sec:conclusioni},
the results are summarised.

Throughout this paper, the abundances are linearly scaled according to
the \citet{grev98} solar composition (i.e., we always mean X/X$_\odot$
for each chemical element X, following the common usage of X-ray based
literature).  We assume 3.63 Mpc as the distance to M82, based on
Cepheids observations by \citet{distanza_m82}.

\section{Observations}
\label{sec:observations}

\begin{figure*}
  \centering
  Please insert 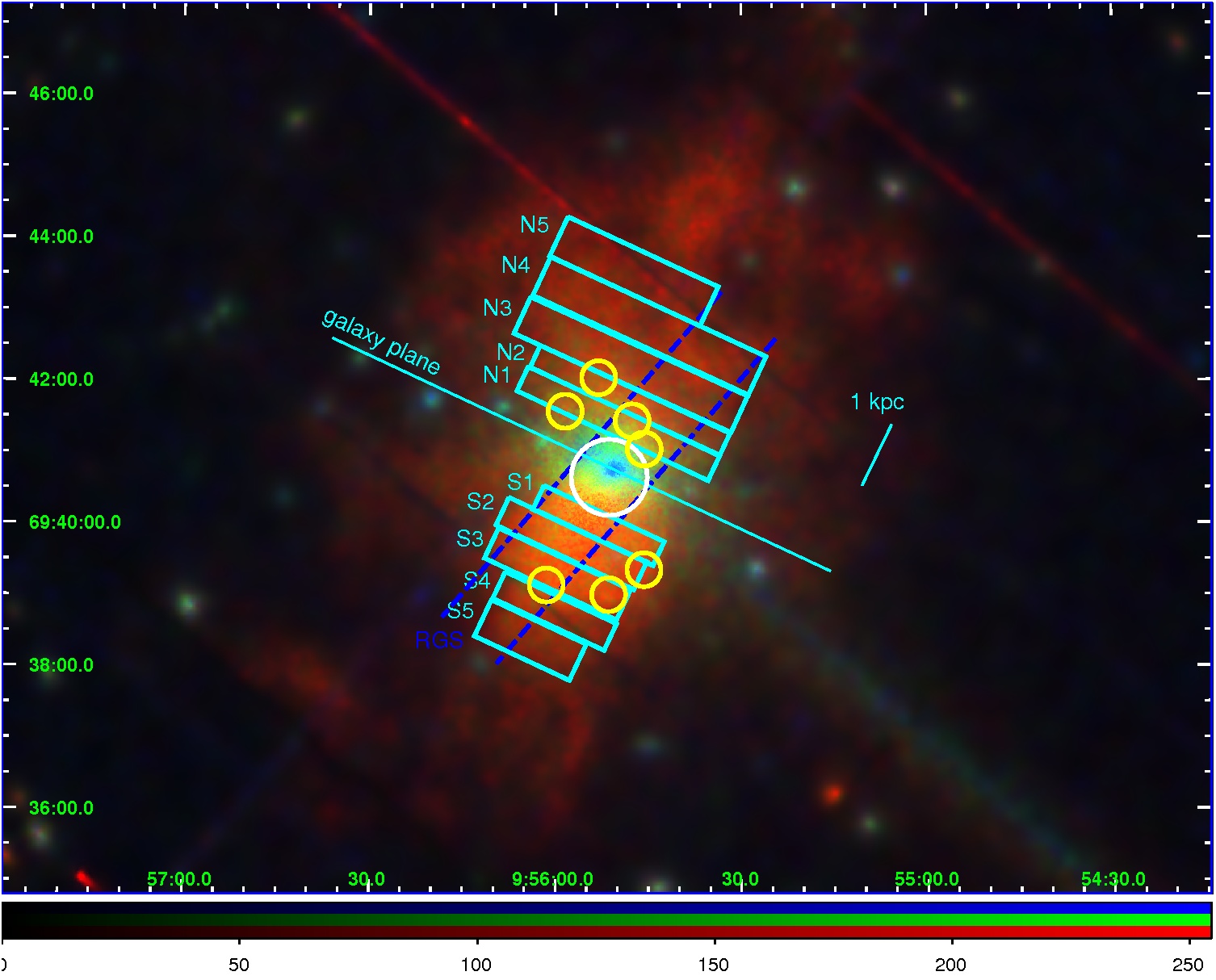 here.
  \caption{Spectral regions superimposed on \xmm\ image. The cyan
    boxes correspond to the outflow regions analysed in
    Sec.~\ref{sec:outflow}, while the small yellow circles show the
    areas around the point sources which have been excluded from the
    extraction of the spectra. The large white circle shows the
    central region whose EPIC spectrum is discussed in
    Sec.~\ref{s:spettri1}; because of the large number of point
    sources found in this region (not evident in the image, and not
    indicated by circles to avoid cluttering the picture) the area
    around them has not been excluded, but rather the point sources
    have been accounted for in the spectral model.  The blue dashed
    lines sketch the RGS spectral extraction region. They are parallel
    to the dispersion direction, and are spaced by $2^\prime$ which
    corresponds roughly to the 90\% of the PSF width, the value used
    in Sect.~\ref{sec:rgs} for the spectral extraction. Along the
    dispersion direction, the PSF is only limited by the instrument
    vignetting.}
  \label{fig:xmmregioni}
\end{figure*}

M82 was observed with \xmm\ on April 21$^{\rm st}$ 2004 for about 100
ks (Obs-id: 0206080101).  After screening for background flares, about
73 ks of data were accepted for analysis.  A true-colour image of M82
is showed in Fig.~\ref{fig:xmmregioni}, with the red, green and blue
colour channels representing the 0.4--1.0 keV, 1.0--2.0 keV and
2.0--8.0 keV bands, respectively. The images in the three bands have
been smoothed with the {\tt asmooth} SAS task, and scaled for the best
visual presentation according to \citet{lupton04}.  Superimposed on
the image, we also show the regions from which the EPIC spectra were
extracted. The RGS spectrum covers an area centred on the galaxy
nucleus and extends approximately over the region shown in
Fig.~\ref{fig:xmmregioni} as the blue dashed parallel lines.

\section{Spectroscopy of the central regions of M82: EPIC data}
\label{s:spettri1}

The EPIC data for both the MOS \citep{XMM-MOS} and {\em pn} \citep{XMM-PN}
cameras were analysed with the SAS (version 6.5.0) and XSPEC (version
11.3.2z) software.  We began the spectral analysis from the central
region of M82, by extracting spectra from a circle with diameter
$1^{\prime}$ and centred on the coordinates 09:55:51 and $+$69:40:39.
In this area the largest emission from hot plasma is present; however,
many point sources exist in the same region \citep{grif00}. Thus, the
spectral model should account for both plasma emission and point
sources, along with considerable absorption.  

\subsection{Spectrum of underlying point sources}
To define an optimal model for the point sources, we analysed the
spectrum in the 3-8 keV band, and found that it can be described by a
power law with photon index $\Gamma=1.60_{-0.03}^{+0.04}$ (all errors
on spectral parameters are at $90\%$), and $\chiq=682$ (with 571
degrees of freedom, hence $\chiqr\sim 1.19$). This emission may be
ascribed, for all practical purposes, to M82 X-1 \citep{matsumoto01}
because the luminosity of this source is much larger than the other
ones \citep{grif00}.  This best-fitting value for $\Gamma$ was not
kept frozen in the spectral fits discussed below, but rather used as a
consistency check.  The point sources, although accounted for in the
spectral fits, will not be discussed further. Results about M82 X-1
from this observation have been presented in
\citet{mucciarelli06}. M82 X-1 is also clearly visible in
Fig.~\ref{fig:xmmregioni} as the blue spot near the galaxy centre.

\subsection{Background spectra for extended sources}
\label{s:background}

The background spectra were derived from blank-sky
observations. Because of the source extent, and the spatial dependence
of the background spectrum on the \xmm\ CCDs, it is in fact impossible
to obtain background fields from the M82 observation. While background
spectra could be extracted near the edges of the field of view where
the M82 emission is negligible, these spectra would be different from
what is expected in the detector region covered by M82.  Thus we used
the \xmm\ blank-sky files provided by the Birmingham group
\citep{birmblanksky}.

Blank-sky observations have the advantage that a background spectrum
can be extracted in the same detector region covered by the source,
but they need greater care in checking that the background level be as
close as possible to the source's one. In order to match the
background brightness, we selected two large circular regions (radius
$\sim 2.35\arcmin$) in a position several arcmin off-axis where no
emission from M82 is present, and extracted spectra from both the M82
data and blank-sky files. In this way, spectra extracted from the same
detector region are compared in order to scale the brightness of the
latter to the former's.  We found only small differences: the
blank-sky/source flux ratios in these two regions are $0.97\pm 0.03$
(MOS1), $0.91_{-0.04}^{+0.03}$ (MOS2) and $0.86_{-0.02}^{+0.03}$ ({\em
  pn}), and are largely independent from the details of model fitting.
Thus, in the following analysis we rescaled every background spectrum
by means of the BACKSCAL keyword in the FITS files.

\subsection{Temperature distribution of the plasma}

It is customary to model the gaseous emission in star forming galaxies
with one or more single-temperature plasmas. A `warm' component around
0.7--1 keV is often reported (e.g.\ \citealt{pta99}). A `hot', high
energy component is also sometimes present \citep{dellaceca97,cappi99}
in studies based on ASCA or BeppoSAX data, described either as a
power-law, or as a thermal spectrum with $kT\sim 4$--10 keV. Because
of the poor spatial resolution of those satellites it is not clear if
this component is really diffuse or rather due to point sources. Thus,
some care should be taken in comparing literature results to this
paper, since in the following we will account for both point sources
(via the power law discussed in the previous paragraph) and for hot
gas. A `cold' component, around 0.1--0.3 keV, has been also
occasionally observed \citep{dellaceca99,tsuru07}.  Moreover, a
potential pitfall has been discovered \citep[the so-called
`Fe-bias',][]{buote98}, which occurs when fitting single-temperature
models to data which are intrinsically multi-temperature, and which
leads to artificially low abundances.

The use of multi-temperature models that allow a fine control of the
Differential Emission Measure (DEM) and are also insensitive to the
Fe-bias might be preferable. However, this method has been seldom used
because of its demanding requirements in terms of data quality and
computing time. The quality of our M82 observation is sufficiently
high to allow for the first time the use of multi-temperature models
even in spatially-resolved
spectroscopy. 

Two spectral codes, MEKAL \citep{mekal} and APEC \citep{apec}, are
currently used in the literature. However, while several versions of
the common MEKAL code are present with multi-temperature flavours
\citep{lemen89,singh96} in the XSPEC distribution, no
multi-temperature APEC model is present.  Since we intended to test
which code offers the best performances, we built multi-temperature
APEC models for XSPEC, which behave like the MEKAL ones, with the only
difference that the APEC routines are called.

We used initially the XSPEC \cSpvmkl\ model, which 
parametrizes the plasma DEM (i.e.\ the amount of plasma
at different temperatures) with a $6^{\rm th}$ order polynomial, and
allows to vary the abundances.

\begin{figure*}
  \centering
  \includegraphics[width=.8\columnwidth]{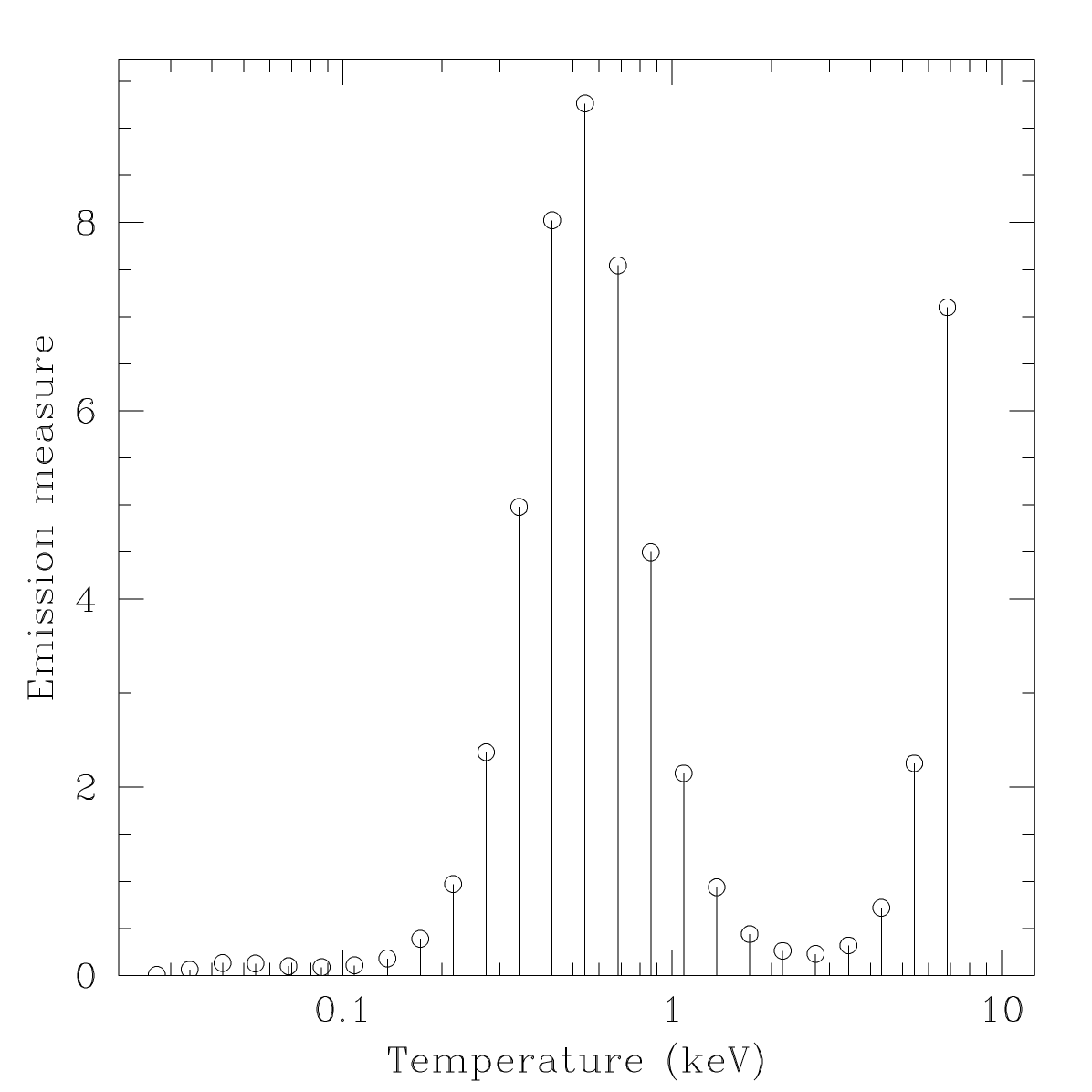}  
  \includegraphics[width=.8\columnwidth]{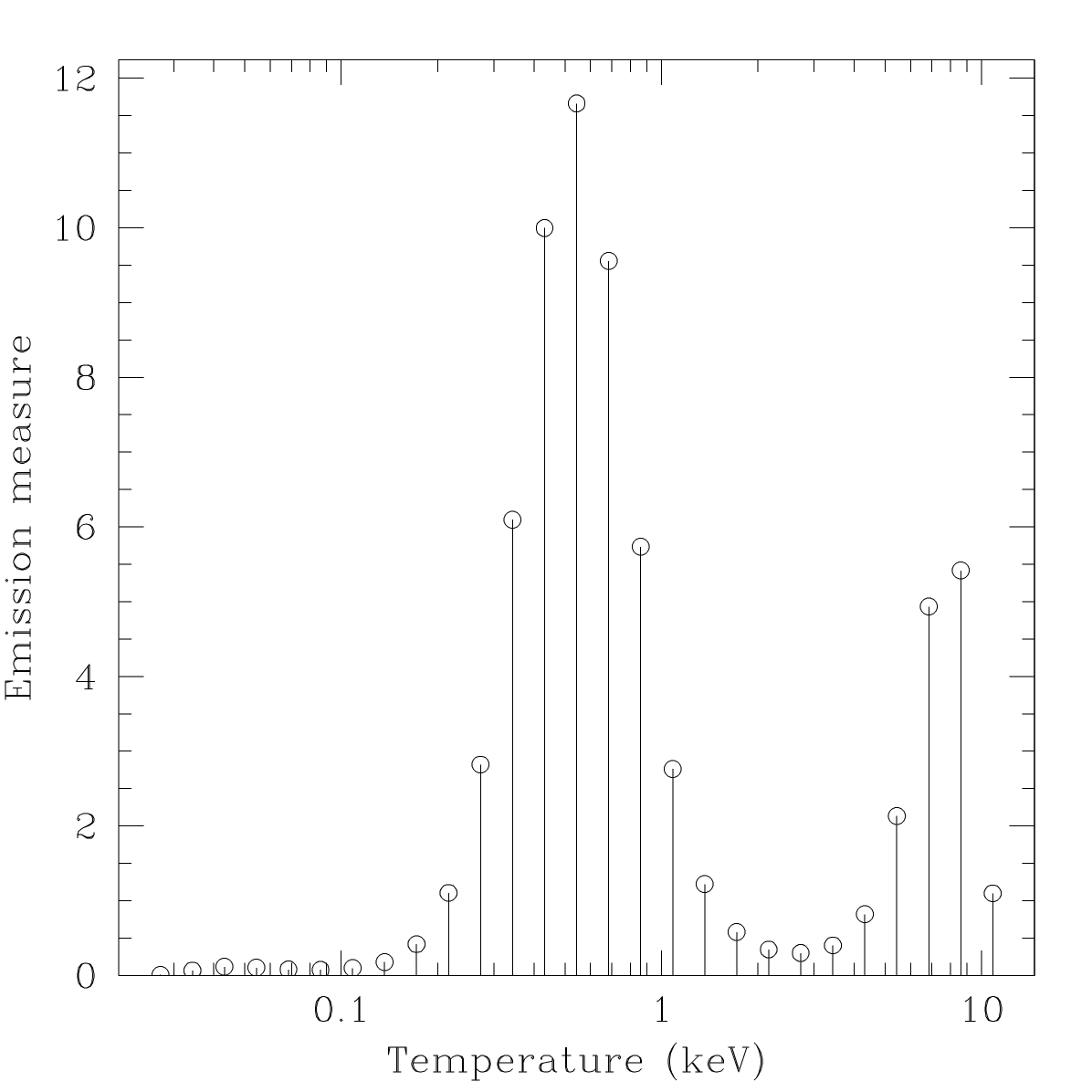}  
  \caption{The best-fitting DEM for the central regions of M82 by using
    the standard {\tt c6pvmkl}\ model (left panel) and its
    enlarged energy array version (right panel). The total spectrum is
    computed by adding several components, one for each sampling point
    shown here.}
  \label{fig:DEM}
\end{figure*}

\begin{figure*}
  \centering
  \includegraphics[angle=-90,width=.9\textwidth]{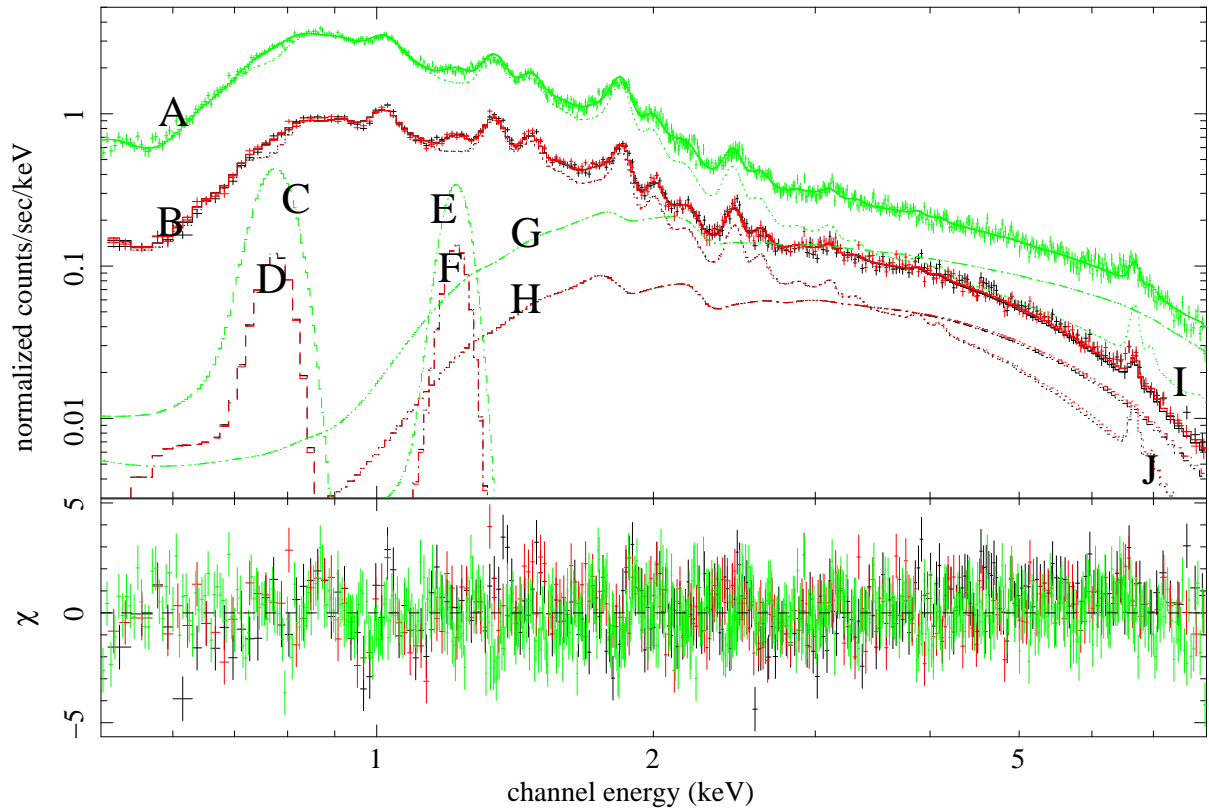}
  \caption{EPIC spectra of the central ($2^\prime$) region of
    M82. Upper panel: spectra with models. Lower panel: residuals,
    plotted in terms of sigmas with error bars of size one.  A:
    observed spectrum (crosses) and model (solid line through the data
    points) convolved with the instrumental response of the {\em pn}.
    B: same, for the MOS1 (black) and MOS2 (red); these two cameras
    give nearly identical results so that the curves and points
    overlap. C and E: Gaussian line components of the model, for the
    {\em pn}.  D and F: same, for the MOS1 and MOS2.  G: absorbed
    power-law component of the model, for the {\em pn}. H: same, for
    the MOS1 and MOS2. I: thermal plasma (MEKAL) component of the
    model, for the {\em pn}. J: same, for the MOS1 and MOS2.  Similar
    results are obtained with the APEC model (see also
    Fig.~\ref{fig:confronto-residui}).}
  \label{fig:epic-mekal-apec}
\end{figure*}

However, this model allows only temperatures with $kT\le 7$ keV,
causing the best-fitting DEM to have a sharp cut, clearly noticeable in
Fig.~\ref{fig:DEM} (left panel).  Feeling that this limitation was
somewhat artificial, we modified the available multi-temperature
plasma models by enlarging the allocated energy range, so that the
high-temperature part of the DEM could be better sampled
(Fig.~\ref{fig:DEM}, right panel).  The improvement in $\chi^2$ was
however not significant, indicating that the DEM sampling was already
sufficiently good: as it may be seen, the high-temperature component
is distributed like a rather narrow bell curve, so little difference
is found between a single temperature element and the sum of a few
elements with very similar temperatures.

\subsection{Results and comparison of plasma codes}

\begin{table*}
\centering
\begin{tabular}{lcccccc}
\hline
    &M. mospn  
    &M. mos
    &M. pn
    &M. mospn no-lines
    &A. mospn
    &A. mospn no-lines\\
\hline
%
%
$N_{H,{\rm th,MOS}}$ ($10^{22}$ cm$^{-2}$) 
          &$0.45\pm 0.01$  &$0.47\pm 0.01$  &---             &$0.45\pm 0.01$ &$0.42\pm 0.01$ &$0.41\pm 0.01$\\
$N_{H,{\rm th,PN}}$ ($10^{22}$ cm$^{-2}$) 
          &$0.44\pm 0.01$  &---             &$0.44\pm 0.01$  &$0.45\pm 0.01$ &$0.41\pm 0.01$ &$0.40\pm 0.01$
\smallskip\\
O         &$0.28\pm 0.02$  &$0.19\pm 0.04$  &$0.29\pm 0.03$  &$0.19\pm 0.02$ &$0.15\pm 0.02$ &$0.13\pm 0.01$\\
Ne[,Na]   &$0.42\pm 0.02$  &$0.49\pm 0.03$  &$0.35\pm 0.03$  &$0.27\pm 0.02$ &$0.46\pm 0.02$ &$0.38\pm 0.02$\\
Mg,Al     &$1.00\pm 0.02$  &$1.01\pm 0.03$  &$0.83\pm 0.02$  &$0.72\pm 0.02$ &$0.77\pm 0.02$ &$0.70\pm 0.02$\\
Si        &$1.05\pm 0.03$  &$1.16\pm 0.04$  &$1.01\pm 0.04$  &$0.93\pm 0.02$ &$1.03\pm 0.03$ &$0.96\pm 0.02$\\
S         &$1.05\pm 0.05$  &$1.29\pm 0.08$  &$0.98\pm 0.08$  &$0.98\pm 0.05$ &$1.18\pm 0.06$ &$1.10\pm 0.06$\\
Ar,Ca     &$1.57\pm 0.30$  &$1.83\pm 0.47$  &$0.85\pm 0.42$  &$0.74\pm 0.25$ &$1.07\pm 0.28$ &$0.48\pm 0.26$\\
Fe,Ni     &$0.33\pm 0.01$  &$0.31\pm 0.01$  &$0.33\pm 0.01$  &$0.28\pm 0.01$ &$0.30\pm 0.01$ &$0.29\pm 0.01$\\
norm$_{\rm th}$ &$2.3\e{-4}$&$2.4\e{-4}$    &$2.4\e{-4}$     &$2.8\e{-4}$    &$2.7\e{-4}$    &$2.6\e{-4}$
\medskip\\                                                           
%
%
$N_{H,{\rm pow}}$ ($10^{22}$ cm$^{-2}$) 
            &$2.26\pm 0.09$ &$5.4\pm 0.5$   &$3.5\pm 0.4$  &$5.25\pm 0.27$ &$4.70\pm 0.25$ &$5.27\pm 0.25$\\
$\Gamma$    &$1.51\pm 0.03$ &$1.48\pm 0.09$ &$1.7\pm 0.1$  &$1.62\pm 0.05$ &$1.59\pm 0.05$ &$1.63\pm 0.05$\\
norm$_{\rm pow}$ &$1.2\e{-3}$  &$8.3\e{-4}$ &$1.2\e{-3}$   &$1.1\e{-3}$    &$1.0\e{-3}$    &$1.2\e{-3}$
\medskip \\
%
%
$\chiq$     &2743 (2267)    &1101 (881)     &1471 (1365)   &3172 (2271)    &2779 (2267)    &2905 (2271) \\
$\chiqr$    &1.21           &1.25           &1.08          &1.40           &1.23           &1.28 
\bigskip \\
%
%
line energy (keV)           &$1.234\pm 0.011$
                                      &$1.240_{-0.13}^{+0.005}$
                                                      &$1.219_{-0.005}^{+0.007}$
                                                                    &---            &$1.222_{-0.002}^{+0.004}$ &---\\
line wavelength (\AA)       &$10.05\pm 0.09$
                                      &$10.00_{-0.04}^{+0.11}$
                                                      &$10.17_{-0.06}^{+0.04}$
                                                                    &---            &$10.15_{-0.04}^{+0.02}$ &---\\
line eq.\ w.\ (eV)          &22.6     &15.8           &15.8         &---            &7.74            &---         
\smallskip\\
line energy (keV)           &$0.777_{-0.006}^{+0.002}$         
                                      &$0.783_{-0.007}^{+0.003}$ 
                                                      &$0.771_{-0.006}^{+0.008}$
                                                                    &---            &$0.767\pm0.006$ &---\\
line wavelength (\AA)       &$15.95_{-0.03}^{+0.13}$
                                      &$15.82_{-0.5}^{+0.14}$
                                                      &$16.08_{-0.16}^{+0.12}$
                                                                    &---            &$16.15\pm0.12$  &---\\
line eq.\ w.\ (eV) &22.4     &31.7           &16.4          &---            &13.6            &---         \\
\hline
\end{tabular}
\caption{Spectral fit parameters for the EPIC data in the inner
  region. All quoted errors are at $90\%$ confidence level.
The rows from top to bottom show: column density of absorbing material
relative to the thermal component in MOS and PN data; chemical
abundances of the thermal component relative to solar values;
normalisation of the thermal component (units as in
Eq.\ref{eq:unitanorm}); column density of the absorbing material
relative to the power-law component; power-law photon index;
normalisation of the power-law component (in units of
photons~s$^{-1}$~cm$^{-2}$~kev$^{-1}$ at 1 keV); $\chiq$ and reduced
$\chiq$; equivalent widths of the CE lines.
The quoted errors are referred to the $90\%$ confidence level, and
have been calculated with all other parameters in the spectral fit
(DEM, abundances, $N_H$, power-law parameters) as frozen.  If this
assumption is relaxed, e.g. by thawing other element's abundances, the
O, Ne and Fe errors relative to the MEKAL-mospn case would be $\pm0.03$,
$\pm0.02$ and $\pm0.01$ respectively.  By relaxing it even more, e.g. by thawing
also the six DEM parameters, one would get $\pm0.05$, $\pm0.05$ and
$\pm 0.03$
respectively.
}
  \label{tab:src0}
%
%

\end{table*}

The abundances for O, Ne, Mg, Si, S, Fe were left free to vary, since
these elements have strong lines in the considered wavelength range.
Other $\alpha$-elements with weaker lines (Na, Al, Ar, Ca), which
would not be sufficiently constrained on their own, were initially
left at the Solar value. However, after noticing that the metals close
in atomic number have similar abundances relative to the solar values
(Table~\ref{tab:src0}), we hooked Na and Al to Mg, and coupled Ar and
Ca together in order to achieve a better $\chiq$. An F-test showed
that this approach, as opposed to leaving Na, Al, Ar and Ca to solar
values, leads to an improvement of the $\chiq$ with significance
$>99.5\%$. While this can be regarded as low significance, we still
preferred this approach rather than assigning arbitrary values. Also,
the abundances for the strong-line elements did not change
significantly.  The best-fitting abundances are all reported in
Table~\ref{tab:src0}.

To account for absorption, we used the T\"{u}bingen model
\citep[\texttt{tbabs},][]{tbabs}, which should be regarded as an
update of the common, and older, Wisconsin model
\citep[\texttt{wabs},][]{morrison83}. A value of $4.0\e{20}$ cm$^{-2}$
was used for the Galactic absorption.

The best-fitting values for the slope of the power-law component lie in
the range 1.5--1.7, depending on the model (Table~\ref{tab:src0}),
consistently with emission from X-ray binaries.

The best-fitting $\chiq$ depends on the instrument whose data are
analysed, on the plasma code (MEKAL or APEC), and on whether two lines
possibly due to charge exchange (CE) emission (see
Sect.~\ref{sec:CE-EPIC}) are modelled.  Not including the CE lines in
the model, a switch from the Mekal- to the APEC-based model improves
the fit: for the Mekal one, one gets, considering MOS+{\em pn}
spectra, $\chiq=3172$ with 2271 degrees of freedom (d.o.f.;
$\chiqr=1.40$); for the APEC one, $\chiq=2905$ ($\chiqr=1.28$). A
visual inspection of the residuals confirms that the APEC model
provides a slightly better representation of the data. On the other
hand, after including the CE lines, the difference in $\chiq$ greatly
reduces, and both models yield similar values for the $\chiqr$.  The
equivalent widths of the CE lines differ, between the Mekal and APEC
models, by a factor of $\sim 2$--3. The spectra, with best-fitting models
and $\chiq$ values, are shown in Fig.~\ref{fig:epic-mekal-apec}. A
comparison of the residuals is shown in
Fig.~\ref{fig:confronto-residui}.

\subsection{Charge exchange emission}

\begin{figure*}
  \centering
  \includegraphics[width=.7\textwidth]{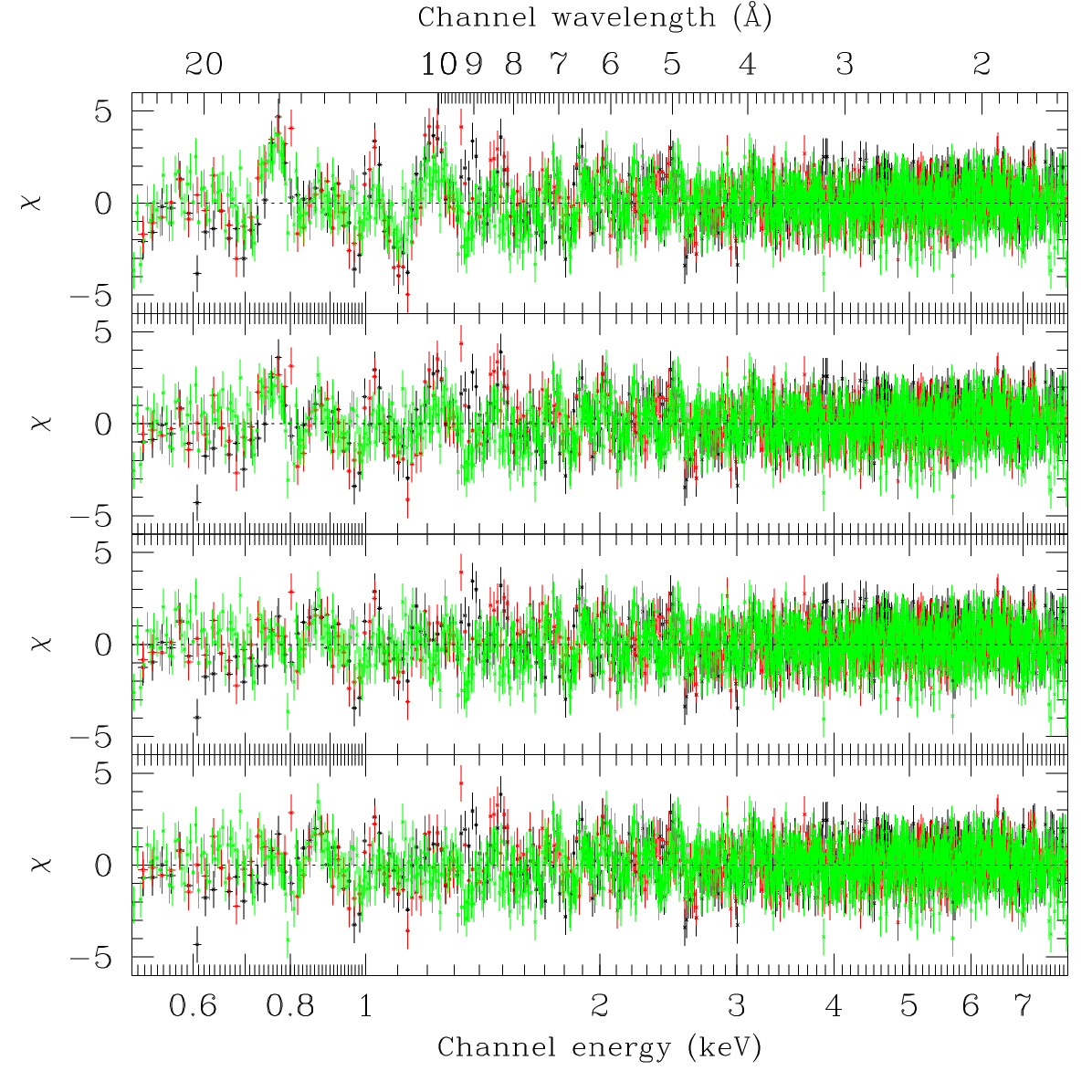}
  \caption{Comparison of residuals of EPIC spectra of the centre of
    M82 from different models. From top to bottom: Mekal without CE
    lines; APEC without CE lines; Mekal with CE lines; APEC with CE
    lines. The $\chiqr$ values are 1.40, 1.28, 1.21 and 1.23,
    respectively (see also Table~\ref{tab:src0}). Black points: MOS1;
    red: MOS2; green: {\em pn}.}
  \label{fig:confronto-residui}
\end{figure*}

Two spectral lines which are not accounted for by the Mekal and APEC
models, showed prominently in the residuals around 16 \AA\ (0.78 keV)
and 10 \AA\ (1.2 keV); they may be best seen in the uppermost panel in
Fig.~\ref{fig:confronto-residui}. The best-fitting wavelengths and
equivalent widts derived by parametrizing the lines with Gaussian
models are reported in Table~\ref{tab:src0}.

\label{sec:simulazioni}
The significance for all these lines when fitting MOS, {\em pn}, and joint
MOS+{\em pn} data can be put at a confidence level $>99.97 \%$. To estimate
the confidence level, we used Montecarlo simulations
\citep{protassov02} which can be briefly outlined as follows:
\begin{enumerate}
\item the model is fit to the data both with and without the line
  component (`line-model' and `null-model', respectively);
\item the variable 
  \begin{equation} \label{eq:F}
F_{\rm data}={\chi^2_{\rm null}-\chi^2_{\rm line}\over
 {\chi^2_{\rm null}\over \rm{dof}_{\rm line}} }
  \end{equation}
\citep{bevington} is calculated for the data;
\item $N$ sets of fake data files and fake backgrounds are generated
according to the null-model.
\item the fake data are fit with both the null-model and the
  line-model, and the $F$ statistic as defined in Eq.~(\ref{eq:F}) is
  calculated for each of them;
\item if $M$ is the number of fake data sets for which $F_{\rm
    fake}>F_{\rm data}$, the confidence level is defined as $1-M/N$.
\end{enumerate}

\label{sec:CE-EPIC}
These lines may be considered real also because: i) the effective
areas in all EPIC cameras in the considered energy range are smooth
functions of the wavelength and do not show any feature around the
considered wavelengths; and ii) no background feature is present near
these wavelengths. Moreover, the count rate of the central region of
M82 is about 100 times larger than the background.

We first looked for identifications in the Chianti database
of line emission from astrophysical plasma \citep{chianti}. The best
candidates for the $\sim 10$ \AA\ line would be an Fe XVII line at 10.1340
\AA\ or a Ne X line at 10.24 \AA, while for the $\sim 16$ \AA\ line the
best candidate might be an Fe XVIII line at 15.83 \AA.  However, the
main problem with this interpretation is that it is not clear why these
lines should not be accounted for by the Mekal and APEC models.

A more plausible hypothesis is that these lines, or at least the $\sim
10$ \AA\ one, are due to charge exchange (CE) emission, occurring at
the interface between the hot wind and clouds of cold neutral gas
within the galaxy itself. In the CE framework, ions from the wind
diffuse through the interface into the cold gas, where electrons are
transferred from the neutrals to the ions. Both the resulting ions can
be highly excited and consequently re-arrange their electrons by
emitting photons in the extreme ultraviolet and X-ray part of the
spectrum.  Only line emission occurs (no continuum), and the photon
emission rate is proportional to the wind ion flux. CE emission has
long been known to occur in (optical) nebular spectra
\citep{chamberlain56} and it has been proposed to account for the
X-ray emission from the comet Hyakutake \citep{cravens97}, the solar
system planets \citep[e.g., Mars:][]{dennerl06}, and it might also
play a non-negligible role in explaining the emission from galactic
winds interacting with dense clouds \citep{lallement04}.

In such a CE scenario, the $\sim 10$ \AA\ line may be identified with
emission from neutral Mg at 9.92 \AA\ (1.25 keV), and this would be
consistent with the presence of olivine and pyroxene grains in the
cold clouds \citep{djuric05}.  The $\sim 16$ \AA\ line, on the other
hand, might be emission from excited O VIII ions also resulting from
CE interactions, where the electron is decaying from the $n=3$ to the
$n=1$ level (but see also Sect~\ref{sec:righe-O}).

\citet{djuric05} suggest that, given the grain composition cited
above, the Mg line should be companied by a Si line at 1.73
keV. The addition of this line to the MOS+{\em pn} spectra is
  significant, at the 99.97\% level, for the APEC model only, and does
  not remarkably change the $\chiqr$. It is, however, not significant
  for the MEKAL model (98.9\% level).  The equivalent width would be
  $\sim 3.9$ eV (APEC) or $\sim 4.6$ eV (MEKAL). These values can be
compared to the equivalent widths for the other two lines, which are
found in the ranges 7--24 eV (see Table~\ref{tab:src0} for the
details).

\section{Large field of view, high resolution spectroscopy: RGS data}
\label{sec:rgs}

The RGS instrument \citep{XMM-RGS} is a slitless grating spectrometer
which allows, thanks to its better energy resolution ($E/\Delta E\sim
300$, while the EPIC {\em pn} camera has $E/\Delta E\sim 7$ in the
overlapping energy range), a more precise identification of the
emission lines and of their properties, including their individual
redshifts and ---at least for broad lines--- profiles.  However, it is
sensitive also to extended emission on scales larger than $\sim
1^\prime$, and M82 was observed with the outflow extent along the RGS
dispersion axis, so that for our purposes the line profiles are
dominated by instrumental broadening. Thus we shall not discuss the
line profiles in the following.

The RGS spectra for the inner region were extracted with the {\tt
  rgsproc} tool, considering only events within $90\%$ of the PSF
width (while the PSF width is actually energy-dependent, this
corresponds roughly to $\sim 1\arcmin$) in the cross-dispersion
direction. On the dispersion direction, the incoming photons are only
limited by the instrument vignetting. The alignment of the RGS
cross-dispersion direction is shown in Fig.~\ref{fig:xmmregioni} as
the blue dashed lines. The background spectra were taken from the
blank sky observations using the {\tt rgsbkgmodel} tool.  Following
the current calibration documentation\footnote{The CAL-TN-0030-3-0
  document, available on the XMM web pages.\label{fn:CAL-TN-0030}},
systematic errors for an amount of $5\%$ in the 9--12 \AA\ range, and
of $8\%$ in the 24.5--31 \AA\ range, have been allowed in the fitting
procedure.

The biggest problem in reducing the RGS data is the resolution
degradation due to source extent.  The {\tt rgsxsrc} convolution model
contained in the XSPEC package may be used to account for the source
extent: it uses an image of the source to derive the line broadening
function, assuming that the image profile is the same at all energies.
Unfortunately, this is not true for M82 mainly because of the heavy
absorption to which its central regions are subject. To mitigate this
problem, we used separate input images for the 6--18 \AA\ and 18--30
\AA\ bands, and for narrower bands when appropriate. A source of
uncertainty is the aperture size used by {\tt rgsxsrc} to derive the
broadening function. While a small aperture may introduce artefacts in
the spectrum due to parts of the source not accounted by {\tt rgsxsrc}
yet contributing to the spectrum, a large aperture may include too
much background thus biasing the source profile. We found that an
aperture of $\sim 4\arcmin$ gives the best trade-off.

In the fitting procedure used for the RGS data, the point sources
contribution to the spectra is taken into account by including the
same power-law component discussed in Sect.~\ref{s:spettri1}, with fixed
parameters.

\subsection{Thermal lines}
\label{sec:termiche}

\begin{figure*}
  \centering
  \includegraphics[width=\textwidth,bb=19 144 591 538,clip]{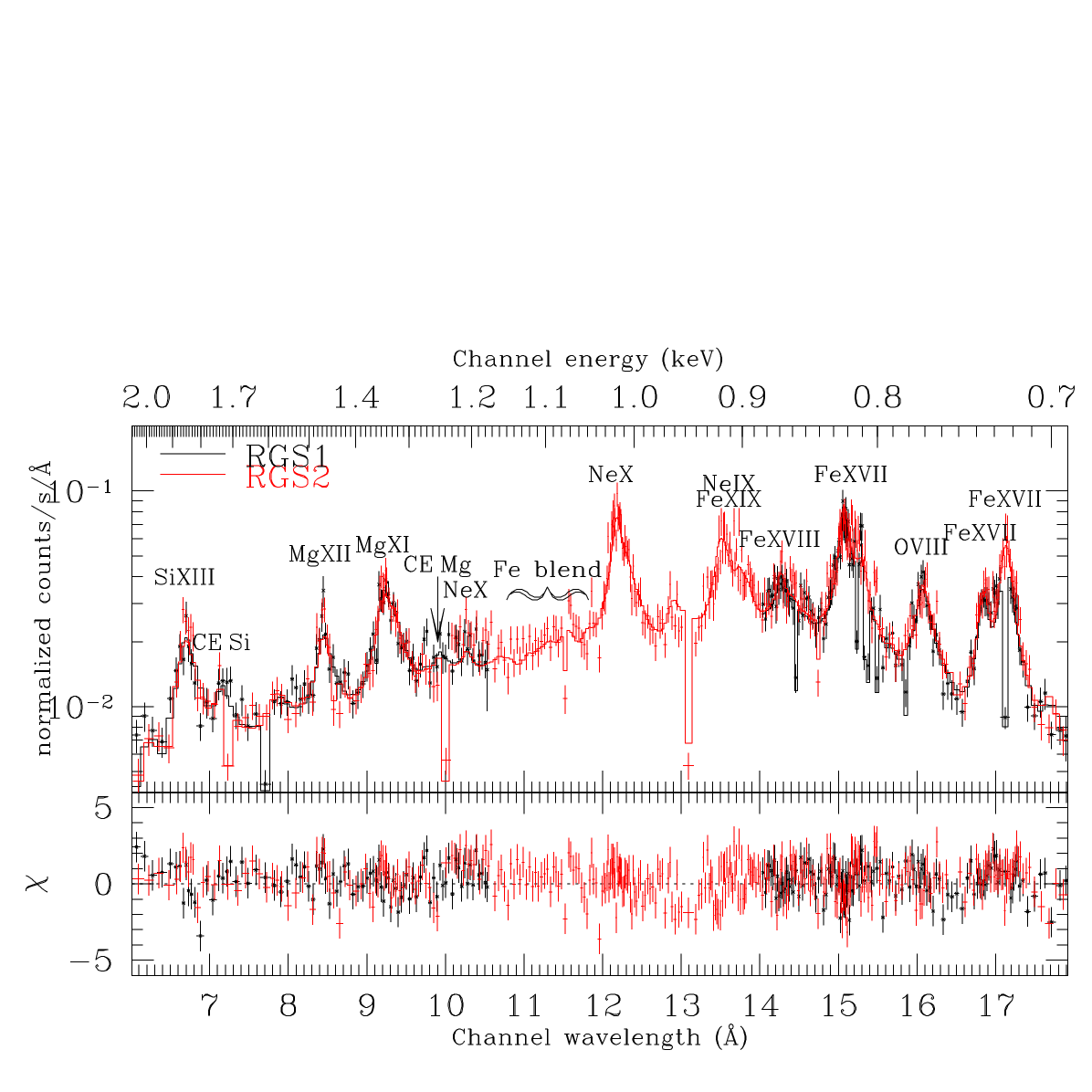}  
  \caption{RGS spectra (upper panel) and residuals (lower panel),
    6--18 \AA. Black data points, with markers: RGS1; red data points,
    without markers: RGS2. The most intense lines are labelled with
    their identifications. In this and in the following pictures about
    RGS data, the points with very low counts looking like wells in
    the spectrum represent the gaps between two consecutive RGS CCDs
    or bad columns.}
  \label{fig:rgs6-18}
\end{figure*}

\begin{figure*}
  \centering
  \includegraphics[width=.3\textwidth,height=.39\textwidth]{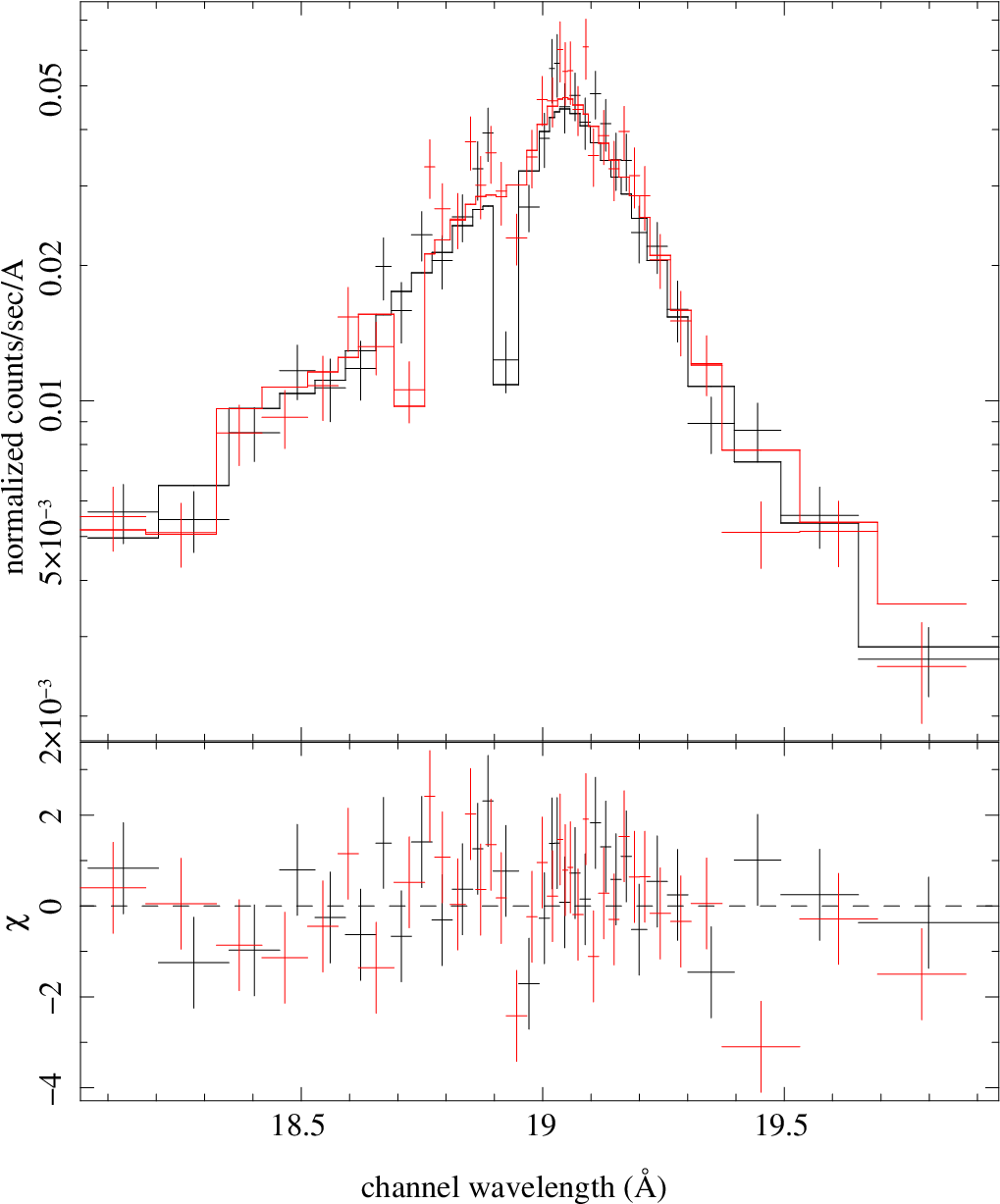}
  \includegraphics[width=.3\textwidth,height=.39\textwidth]{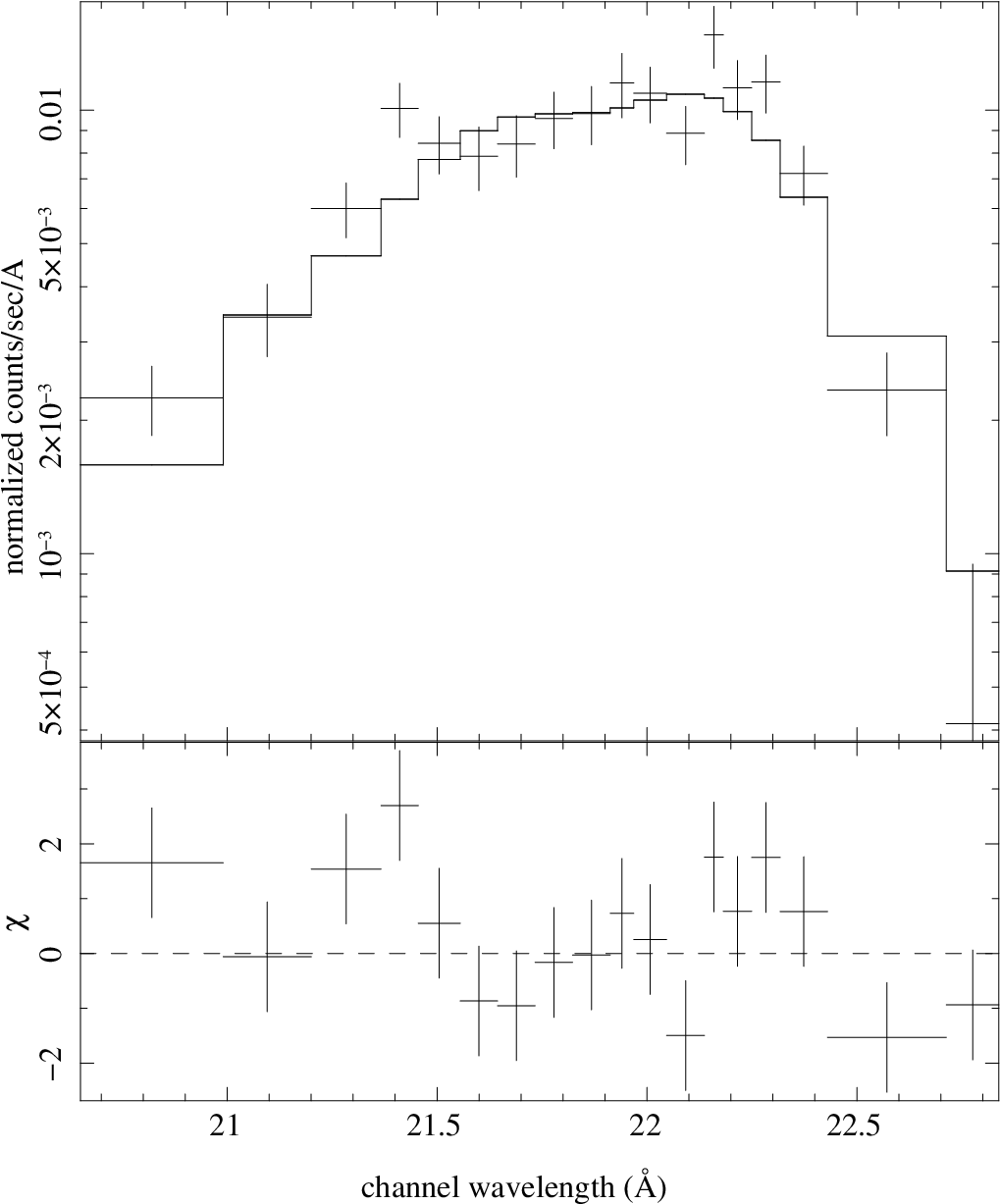}
  \includegraphics[width=.39\textwidth,height=.39\textwidth]{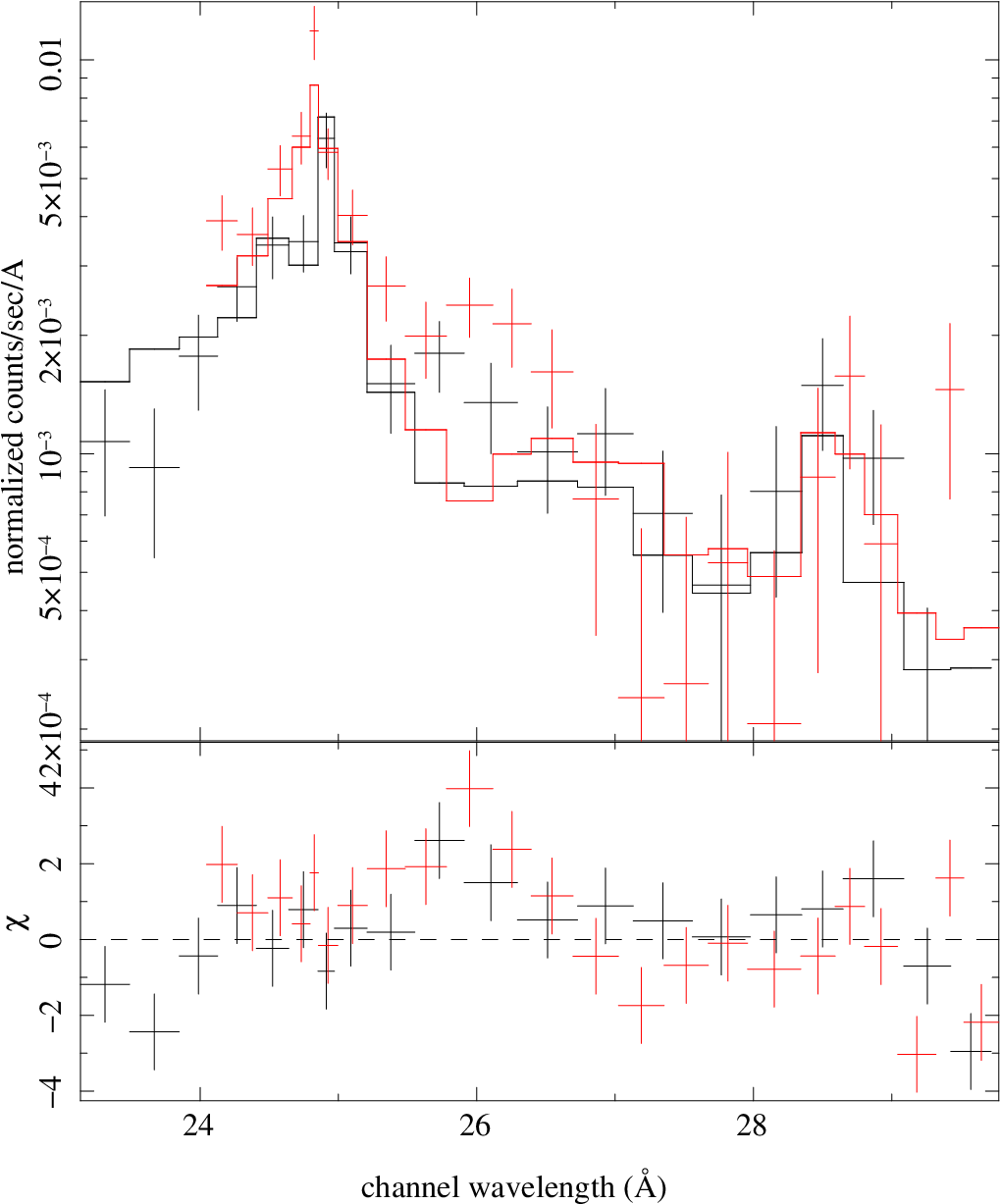}
  \caption{RGS spectra and residuals in the 18--30 \AA\ region.  From
    left to right: spectrum and residuals of the O VIII $\lambda 19$
    \AA\ line, fitted with the APEC model; same, for the O VII triplet
    with best-fitting model; same, for the N VII and C VI lines fitted
    with the APEC model.}
  \label{fig:rgs18-30}
\end{figure*}

The spectra, with the best-fitting model and the residuals, are shown in
Fig.~\ref{fig:rgs6-18} (6--18 \AA) and Fig.~\ref{fig:rgs18-30}
(18--30 \AA), while the spectral results are listed in
Table~\ref{tab:rgs}.  The brightest lines between 6 and 10.5 \AA\ are
from Si XIII (a blended triplet around 6.7 \AA), Mg XII ($\lambda$8.42
\AA), and Mg XI ($\lambda$9.17 \AA).  Other elements with fainter
emission in this region are Al, Na and Ne, while the 10.5--12.0 \AA\
region is dominated by L-shell Fe emission, mainly from Fe XII, Fe
XIII and Fe XIV. The Ne X $\lambda 12.1$ \AA\ line is also prominent.
At longer wavelengths many lines are present from Fe and Ni, plus some
bright lines from Ne and O; the brightest ones are Fe XVII $\lambda
15.0$, 16.8--17.1 \AA; Fe XVIII $\lambda 14.2, 14,4$ \AA; Fe XIX
$\lambda 13.5$ \AA), and from Ne IX ($\lambda 13.4$ \AA) and O VIII
($\lambda 16.0$ \AA).  Longwards of 18 \AA, the lines O VIII $\lambda
19$ \AA, O VII $\lambda 21.6$--22.1 \AA, N VII $\lambda 24.8$ \AA\ and
a hint of C VI $\lambda 28.5$ \AA\ are detected.  Line identifications
and wavelengths are from CHIANTI \citep[and references
therein]{chianti}.

The O lines, because of the possible contribution from the CE
mechanism and because a different best-fitting redshift is needed for
O VIII $\lambda 19$ \AA\ than for the lines at shorter wavelengths,
will be discussed in Sects.~\ref{sec:righe-O}.  Thus, for the
dependence on energy of the line profiles and also because the N VII
and C VI lines have much fewer counts than all other lines, it is best
to restrict the interval to analyse to the 6--18 \AA\ region.  This
restriction does not affect most elements (i.e.\ Ne, Mg, Al, Si, Fe
and Ni) whose lines are weak, if present at all, in the 18--30 \AA\
region.

The best-fitting values for the abundances provided by the two codes are
in agreement within errors for all elements.
The APEC model gives the best-fitting to the observed 6--18 \AA\ spectrum,
with $\chiqr\sim 1.15$.  It was not possible to get a good fit by
using the MEKAL model (the best one has $\chiqr\sim 1.38$).  The
best-fitting abundances from both models are shown in Table~\ref{tab:rgs},
and the APEC-derived ones are also displayed as the red horizontal
lines in Fig.~\ref{fig:logelem}).

The restricted wavelength interval 12-18 \AA\ looks like the best one
in which to test our assumption that the Fe and Ni have equal
abundances, both because of fit goodness ($\chiq\sim 1.05$) and of the
relative paucity of lines from other elements. By using an F-test, the
improvement in $\chiqr$ resulting from thawing the Ni abundances was
not significant (improvement only at 75\% confidence level);
this was achieved by keeping all other parameters except Fe frozen;
thawing O and Ne further reduced the significance of the $\chiq$
improvement.

\begin{figure*}
  \centering
  \includegraphics[height=.47\textwidth,angle=-90]{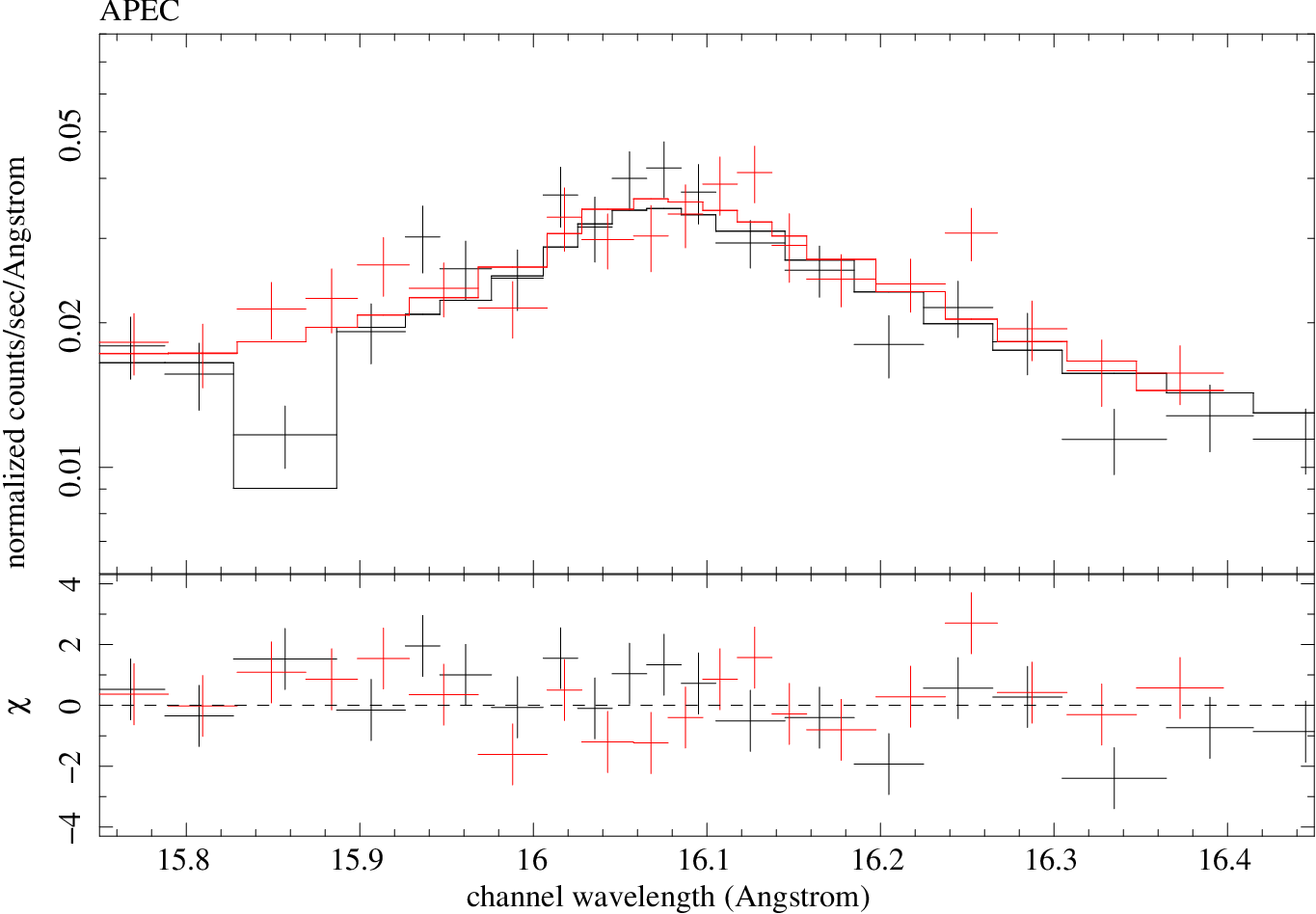}
  \includegraphics[height=.47\textwidth,angle=-90]{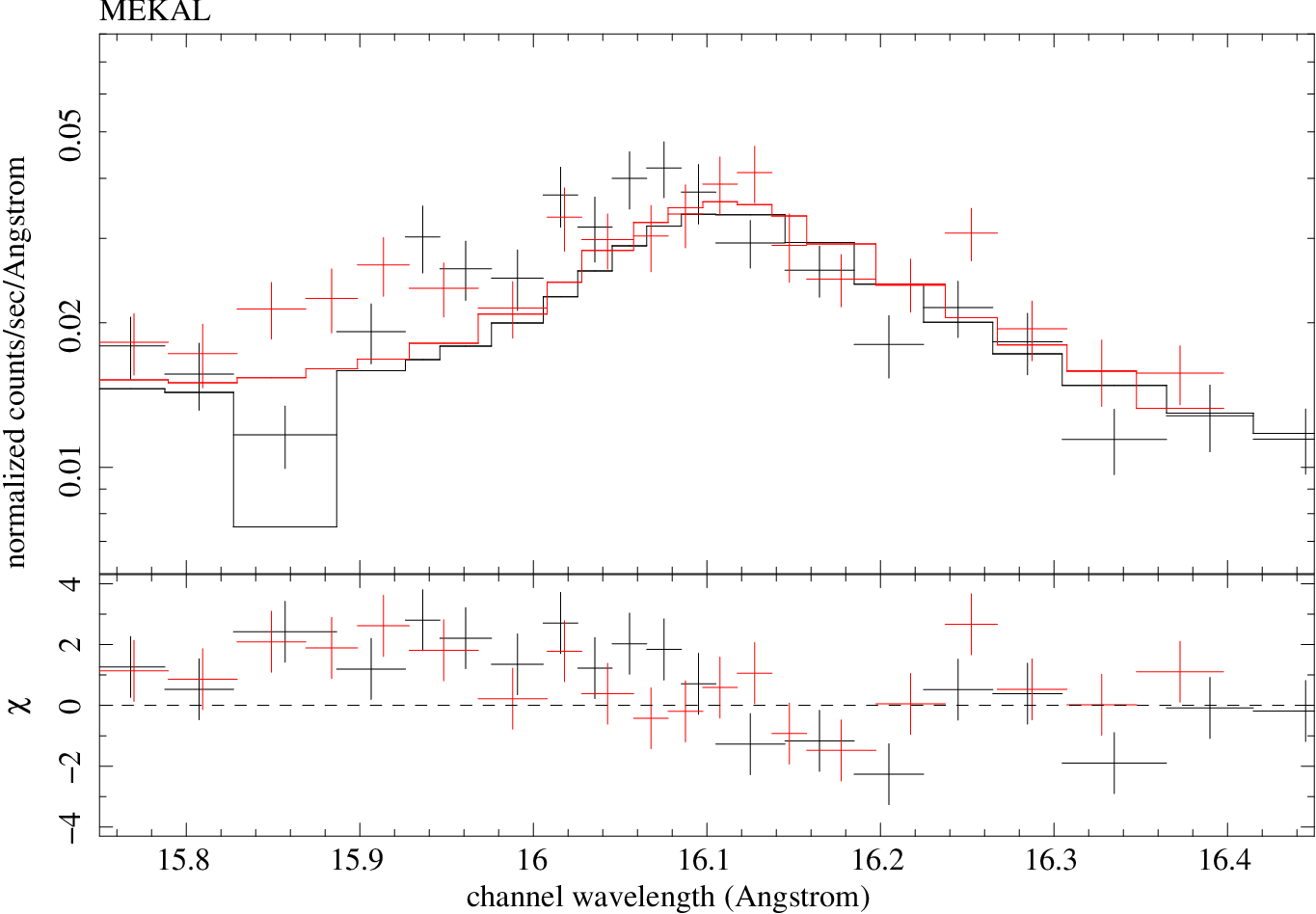}
  \caption{Comparison of the APEC (left) and MEKAL (right) modellings of
    the O VIII $\lambda 16$ \AA\ line.  The goodness-of-fit is
    $\chiqr=1.0$ for the left panel and $\chiqr=1.6$ for the right
    one.  A different modelling of the Fe XVIII lines at 15.4 \AA,
    15.6 \AA\ and 15.8 \AA\ is probably the origin of the
    discrepancies. Crosses: data points. Solid lines: models. Black:
    RGS1. Red: RGS2.}
  \label{fig:OVIII16A}
\end{figure*}

\subsubsection*{The N VII and C VI  lines}
\label{ssec:NVII-CVI}

Because of the strong absorption, few photon counts are available in
the 24--30 \AA\ region where the N VII and C VI lines are present.  No
significant constraint may be put on the C abundance, which would
mainly be obtained from the C VI line at 28.46 \AA\ where the
absorption is maximum. An estimate of the N abundance may be placed at
values larger than solar, albeit with rather large errors.  The random
errors are reported in Table~\ref{tab:rgs}.  As to the systematic
ones, one could make the following consideration. The best-fitting
abundance of N in Table~\ref{tab:rgs} was calculated by using an input
image for {\tt rgsxsrc} in the 24--26 \AA\ band, and the redshift as a
free parameter. Had we included the 24--26 \AA\ region along the 6--18
\AA\ one, with the same 6--18 \AA\ input image for {\tt rgsxsrc}, and
assigned to the N VII line the same redshift of the other considered
ones, then the best-fitting N abundance would have been $1.0\pm 0.2$.

\begin{table}
\centering
\begin{tabular}{lcc}
\hline
    &Mekal       &Apec  
\\
\hline
%
$N_{H}$ ($10^{22}$ cm$^{-2}$)  
                &$0.21\pm 0.01$            &$0.15\pm 0.01$  \\
N               &$5.9_{-2.2}^{+3.4}$          &$3.2\pm 1.6$    \\
O (16\AA)       &$1.31\pm 0.16$            &$0.62\pm 0.10$ \\
Ne[,Na]         &$0.96\pm 0.06$            &$0.95\pm 0.07$  \\
Mg              &$1.69\pm 0.12$            &$1.68\pm 0.10$  \\
Al              &$1.9\pm 1.3$              &$3.1\pm 1.2$    \\
Si              &$2.49\pm 0.23$            &$2.64\pm 0.23 $ \\
Fe,Ni           &$0.51\pm 0.02$            &$0.55\pm 0.02$\\ 
norm$_{\rm th}$ &$2.28\e{-4}$              &$2.15\e{-4}$   
\smallskip\\                                                           


%
%
$\chiq$           &1559.7 (1133)           &1303.4 (1133)    \\ 
$\chiqr$          &1.38                    &1.15
 \\
\hline

%
%
O (19\AA)         &$0.87\pm 0.03$            &$1.41\pm 0.05$ \\
$\chiq$           &284.1 (148)               &181.3 (148)    \\ 
$\chiqr$          &1.9                       &1.22
\\
\hline
%
%
eq.\ w.\ of Mg line (eV) &13.1   &12.4   \\
eq.\ w.\ of Si line (eV) &43.7   &43.3   \\
\hline
\end{tabular}
\caption{Chemical abundances from spectral fits to RGS data.  The
  quoted errors are referred to the $90\%$ confidence level, and have
  been calculated with all other parameters in the spectral fit (DEM,
  abundances, $N_H$, power-law parameters) as frozen.    }
  \label{tab:rgs}
\end{table}

%
%
%
%
%
%

\subsection{The O lines}
\label{sec:righe-O}

Two lines from O VIII ($\lambda 16$ \AA\ and $\lambda 19$ \AA) and one
triplet from O VII (21.6--22.1 \AA) are present in the RGS spectrum.
It is not possible to fit all lines together, either with the APEC or
with the MEKAL models, because these models assume a common redshift
for all lines, which is not the case for O. Thus these lines will be
analysed separately.

An additional complication may be the possible presence of CE
emission, that should company the Mg and Si emission if the ions
impacting on the dust contain O. The relative strengths of the three
components of the O triplet arising from CE have recently been
measured for the first time in laboratory experiments
\citep{beiersdorfer03}, finding that the dominant contribution comes
from the forbidden line. This effect has also been observed
astronomically in Jupiter's auroras, albeit with large errors on the
line fluxes \citep{branduardi07}.  Experiments on CE emission from the
Fe XXV triplet have also confirmed the predominance of the
intercombination and forbidden lines relative to the resonance one
\citep{wargelin05}.

\subsubsection{The O VIII $\lambda 16$ \AA\ line}
\label{ssec:OVIII16A}
The O VIII $\lambda 16$ \AA\ line can be fit together the other lines
in the 6--18 \AA\ range; the abundances derived from this line are
shown in Table~\ref{tab:rgs} along with the results from the other
elements.
A juxtaposition of close-up images of the O VIII $\lambda 16$ \AA\
with both the APEC and MEKAL models is shown in
Fig.~\ref{fig:OVIII16A}. The goodness's of fit in the 15.75--16.45
\AA\ interval are $\chiqr=1.0$ and $\chiqr=1.6$ for the APEC and MEKAL
models, respectively.  From the comparison, one may attribute the
deviations to a different modelling of Fe XVIII lines at 15.4 \AA,
15.6 \AA\ and 15.8 \AA. The APEC-derived value for the O abundance
seems therefore more robust.

\subsubsection*{The O VIII $\lambda 19$ \AA\ line}
\label{ssec:OVIII}

The O VIII line $\lambda 19$ \AA\ formally needs a different redshift
than the lines at shorter wavelength to be fitted. The best-fitting
redshifts are $9.6\e{-4}$ and 0 for the APEC and MEKAL models,
respectively. However, the relative shifts between O VIII $\lambda 19$
\AA\ and the lines at shorter wavelengths are of the same magnitude of
the systematic uncertainties (see Sect.~\ref{sec:zetazeta}). The
uncertainties are mainly due to the dependency of the line broadening
function on the energy. The amount of absorption makes O lines
especially sensitive to this effect, so that using a narrow band image
as the input for {\tt rgsxsrc} gives the best results.

The O abundances extracted from this line, with all other parameters
kept the same as for the 6--18 \AA\ spectrum, are shown in
Table~\ref{tab:rgs}.  While the APEC model gives an acceptable fit
($\chiqr\sim 1.2$) by using the same parameters for the 6--18 \AA\
spectrum, the MEKAL model needs a different value for the column
density or for the normalisation to achieve a good fit.  For the MEKAL
model, had we thawed the column density, we would have got $N_{\rm
  H}\sim(4.5\pm0.1)\e{21}$ cm$^{-2}$, O/O$_\odot\sim 3.6\pm0.1$ and
$\chiqr\sim 1.2$. Had we thawed the normalisation, a similarly high
abundance of O would have been obtained.

\subsubsection*{The O VII triplet}
\label{ssec:OVII}

In the 20.5--23.0 \AA\ the only strong lines are the He-like O VII
triplet, at 21.6 \AA, 21.8 \AA\ and 22.1 \AA, known as the {\em
  resonance}, {\em intercombination} and {\em forbidden} lines,
respectively. We shall retain this nomenclature hereafter.  One
difficulty in fitting the O VII triplet is that all the lines are
blended.  

The RGS1 spectrum in the O VII region (RGS2 is not working at
these wavelengths) is reproduced neither by the APEC nor by the MEKAL
model. Thus in the following we will try and fit the spectrum with
Gaussian lines.
The following line ratios are used in the literature, which have been
shown to be sensitive to the electron density and temperature
\citep{gabriel72}:
\begin{equation}
  \label{eq:erre}
  R(n_{\rm e}) = f/i
\end{equation}
\begin{equation}
  \label{eq:gi}
  G(T_{\rm e}) = (f+i)/r
\end{equation}
where $f$, $i$, and $r$ represent the fluxes in the forbidden,
intercombination and resonant line respectively. 

From \citet{beiersdorfer03} one would get, for CE emission, $R\sim 3$
and $G\sim 2.2$; \citet{branduardi07} gives values which are
consistent, within errors, with \citeauthor{beiersdorfer03}'s results.

For a thermal plasma dominated by collisional ionisation (CI), the
ranges for these parameters are $2\e{-2}\lesssim R\lesssim 4$
(considering densities ranging from $10^{12}$ cm$^{-3}$ to 1
cm$^{-3}$, respectively) and $0.6\lesssim G\lesssim 1.2$ (temperatures
from $\sim 0.35$ keV to $\sim 0.1$ keV, respectively%
\footnote{While temperatures warmer than 0.35 keV may also be
  plausible, we did not consider this possibility in the fit, since
  for higher $T_{\rm e}$ the flux in the whole O VII triplet
  decreases.}%
); however the presence of photo-ionisation may lower $R$ mimicking a
high density plasma; for a hybrid (collisional+photo) plasma one
should consider $2\e{-4}\lesssim R\lesssim 4$ \citep{porquet01}.  Thus
it may be difficult to distinguish between a high density, low
temperature collisional plasma and CE emission. In the following, we
always refer to \citet{porquet01} for the links among $R$, $G$ and the
physical parameters of the plasma.

\begin{figure}
  \centering
  \includegraphics[width=\columnwidth]{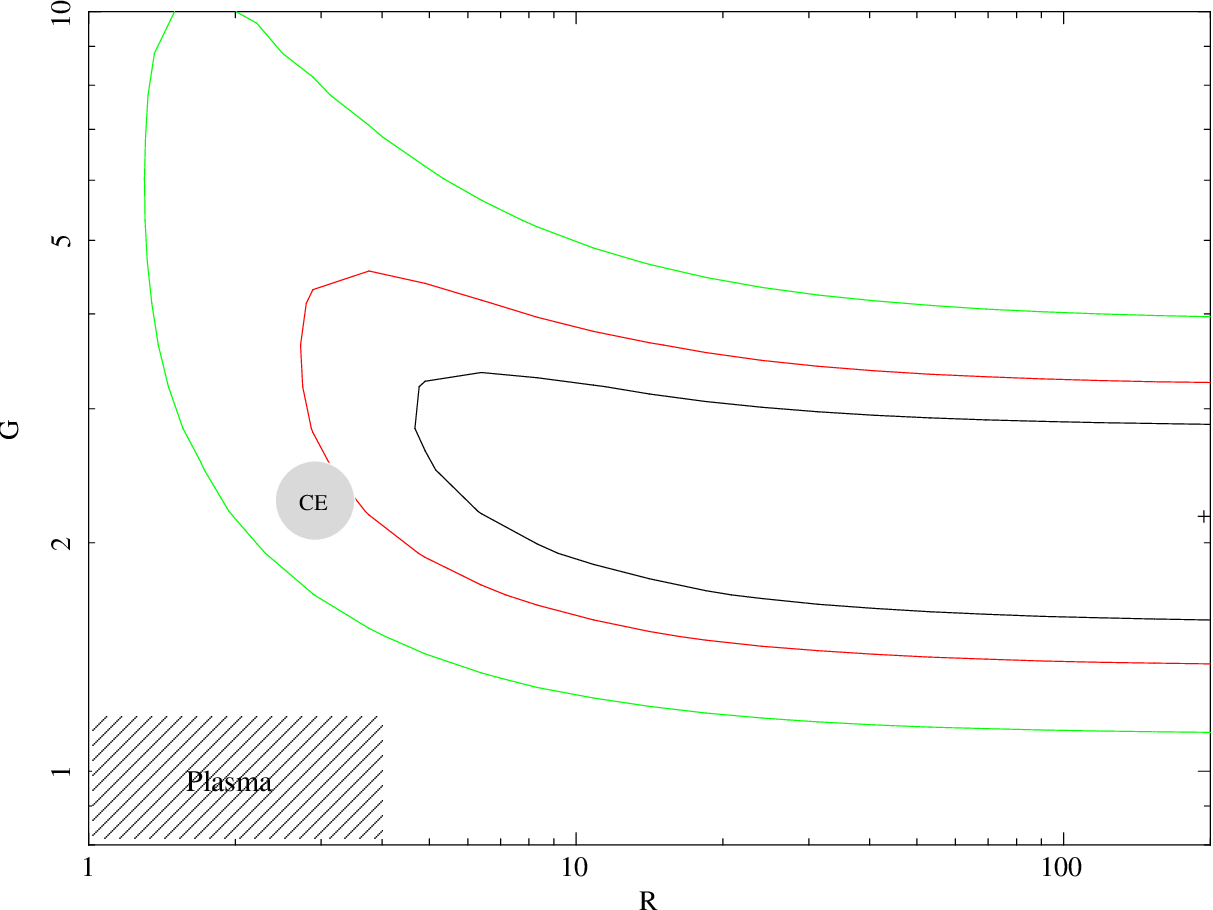}
  \caption{Confidence contours (68.3\%, 90\% and 99\%) for the $R$ and
    $G$ parameters (which describe the line intensity ratios) for the
    O VII triplet. The spectrum is marginally consistent with CE
    emission (shown as the grey circle; the size of the circle is not
    representative of the uncertainties), but is not consistent with
    plasma emission (hashed rectangle).}
  \label{fig:Ovii-contorni}
\end{figure}

It is possible to obtain an acceptable ($\chiqr\sim 1.2$) fit to the
spectrum by using a single triplet of lines, provided that some
additional broadening is present beyond the ones accounted for by the
instrumental response and {\tt rgsxsrc}; this may be parametrized as a
dispersion $\sigma\sim 3.7\pm 1.3$ eV for each Gaussian component of
the triplet. If this broadening were of thermal origin, the
corresponding velocity would be $\sim 8$ km s$^{-1}$. The line
intensity ratios are marginally consistent with CE but not consistent
with plasma emission, as may be seen in the confidence contours
plotted in Fig.~\ref{fig:Ovii-contorni}.  It is not clear why only O
VII needs additional broadening, with respect to all other lines;
while it may be that multiple components (for instance, CE and plasma)
are present at different velocities, the data resolution is not high
enough to disentangle this issue.

\subsection{Lines with possible contribution from Charge-Exchange:
  Mg, Si}
\label{sec:RGS-MgSi}

Some residuals around 7.2 \AA\ could be identified with a CE line from
neutral Si, with a best-fitting energy within errors from what expected
(rest frame wavelength: 7.1280 \AA, or 1.7394 keV,
\citealt{djuric05}). The significance of this line can be estimated,
with the Montecarlo simulations described in
Sect.~\ref{sec:simulazioni} and using the APEC model, at a confidence
level $\ge 99.99\%$.

Another possible identification for this line might be Al XIII
$\lambda$7.17 \AA, which however is not allowed by the APEC model
because an enhancement of this line due to a larger Al abundance
should be companied by a similar increase in the Al XII triplet at
7.76--7.87 \AA, which is not observed.  In order to reproduce the Al
XIII/Al XII intensity ratio, the plasma temperature should be $kT\sim
2$ keV, which is at odds with the DEM structure derived from the other
elements.

The Mg CE line (referred to as `the $\sim 10$ \AA' line in
Sect.~\ref{sec:CE-EPIC}) is detected only at a lower confidence level
(99.8\%).

\begin{figure*}
  \centering
  \includegraphics[width=.8\textwidth,bb=0 148 591 464,clip]{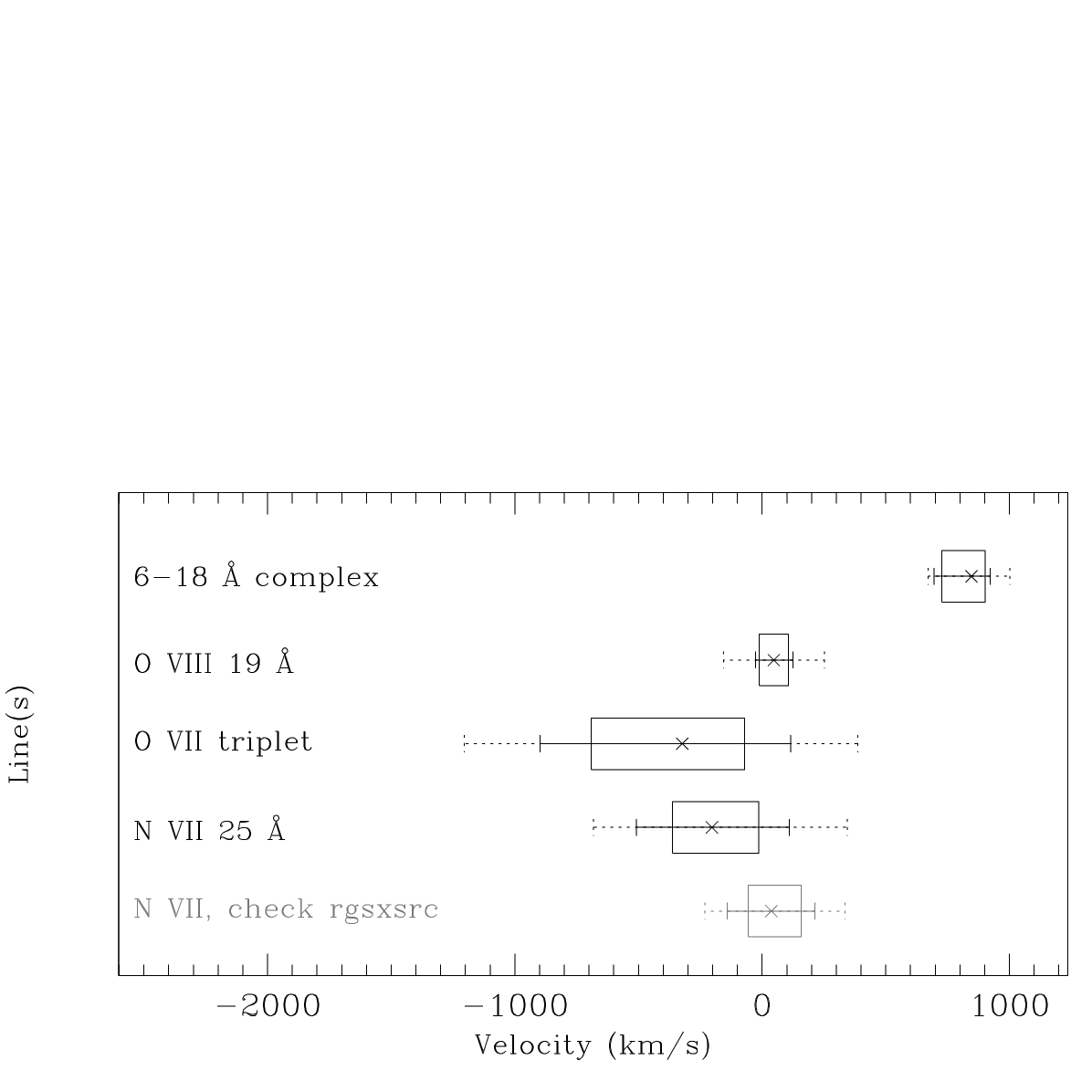}
  \caption{Comparison of best-fitting velocities, relative to the bulk
    motion of M82 ($z=6.77\e{-4}$, \citealt{rc3}), for different lines
    and wavelength intervals in the RGS spectra.  The crosses show the
    best-fitting values, while the boxes represent the 68.3\%
    confidence interval, and the solid- and dotted-line error bars
    show the 90\% and 99\% confidence intervals respectively.  The
    grey whisker labelled `check rgsxsrc' shows the maximum possible
    systematic error arising from an incorrect image supplied to {\tt
      rgsxsrc}, estimated for the N VII $\lambda 25$ \AA\ line using a
    6--12 \AA\ image.  }
  \label{fig:zbaffi}
\end{figure*}

\subsection{Comparison of best-fitting velocities}
\label{sec:zetazeta}
In Fig.~\ref{fig:zbaffi} we show a comparison of the different ranges
for the best-fitting velocities of the X-ray plasma in the 6--18 \AA\
range, the O VIII $\lambda 19$ \AA\ line, the O VII $\lambda
21.6$--22.1 \AA\ triplet, and the N VII $\lambda 25$ \AA\ line. The
lines in the 6--18 \AA\ interval formally have a larger velocity than
the others at longer wavelengths.

In the following we discuss possible systematic uncertainties which
may be responsible of the observed discrepancy between `short-' and
`long'-wavelength lines.
\begin{itemize}
\item We recalculated the $\chiqr$ distribution for the redshift of
  the N VII line using a 6--18 \AA\ image, in order to maximize the
  distance between the wavelengths of the line and of the reference
  image. The result is shown in grey in Fig.~\ref{fig:zbaffi} with the
  label `N VII check rgsxsrc'.  We found that in this case the
  redshift of the N VII line becomes slightly larger, but it is still
  not consistent with the redshift of the lines in the 6--18 \AA\
  range.
\item Using a 6-12 \AA\ image as input for {\tt rgsxsrc} while
  studying the O VIII $\lambda 19$ \AA\ line results in a much worse
  fit ($\chiqr\sim 1.4$) than using an image in the proper band
  ($\chiqr\sim 1.0$). Allowing additional broadening does not make the
  fit better.
\item At variance, splitting the 6--18 \AA\ range in two smaller
  intervals (6--12 \AA\ and 12--18 \AA) did not show any difference in
  the best-fitting velocities, with both cases confirming the 6--18
  \AA\ value.
\end{itemize}

The shift between the `long-' and `short'-wavelength lines amounts to
$\Delta z \sim 3\e{-3}$; considering an average wavelength of 15 \AA,
this corresponds to a $\Delta\lambda=\lambda\, \Delta z\sim 45$ m\AA,
hence to a spatial separation of $\sim 20\arcsec$, if due to an
angular offset of the two emitting regions within the RGS aperture.
This distance is shorter than the aperture considered for {\tt
  rgsxsrc}, and is comparable to the size of the region where there is
the maximum foreground absorption.  Thus the velocity shift may well
be possibly attributed to the apparent spatial separation of the
regions emitting at long and short wavelengths.

\section{Spectroscopy of the outflow}
\label{sec:outflow}

\begin{figure*}
  \centering
  \includegraphics[width=.25\textwidth,angle=-90]{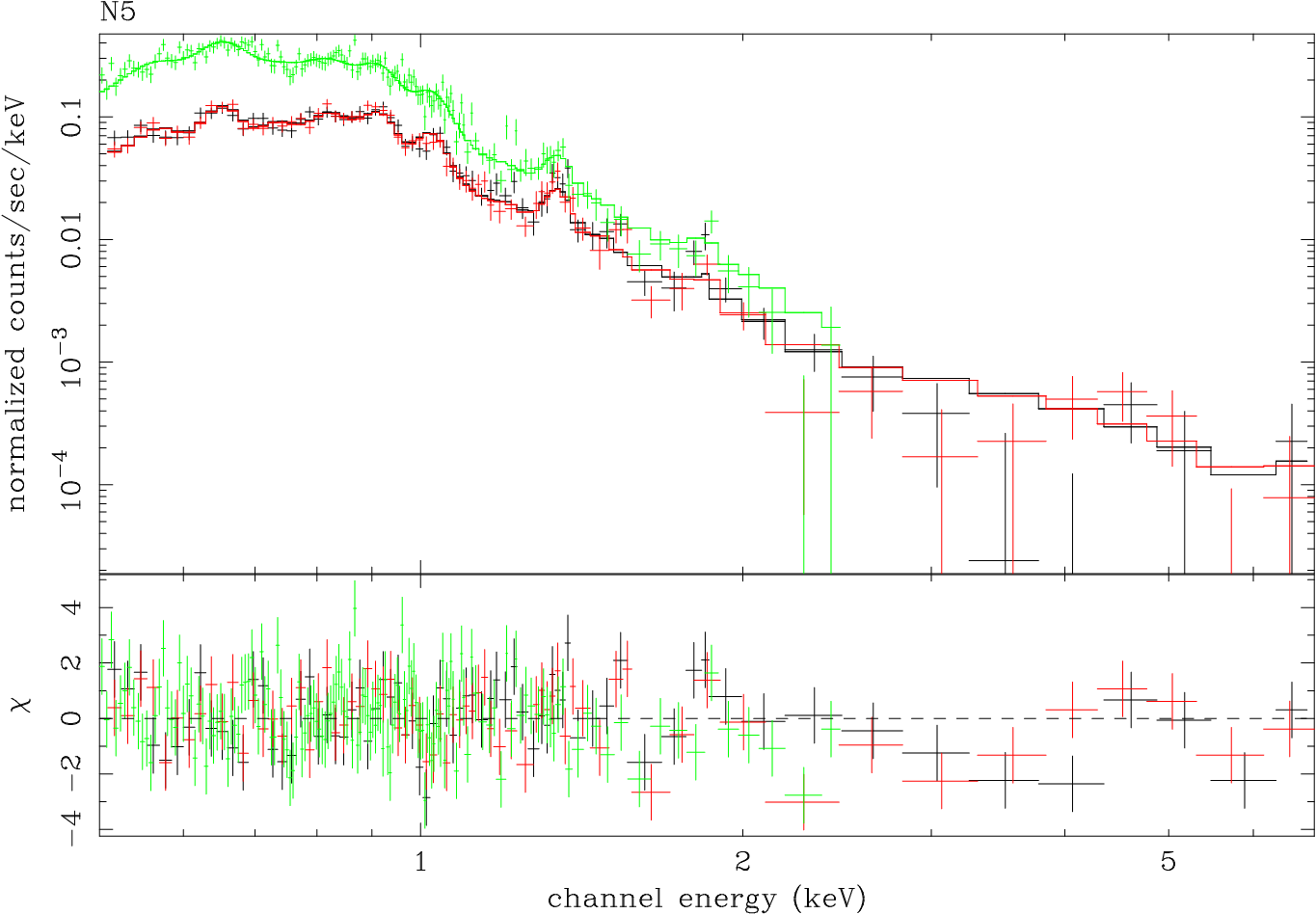}
  \includegraphics[width=.25\textwidth,angle=-90]{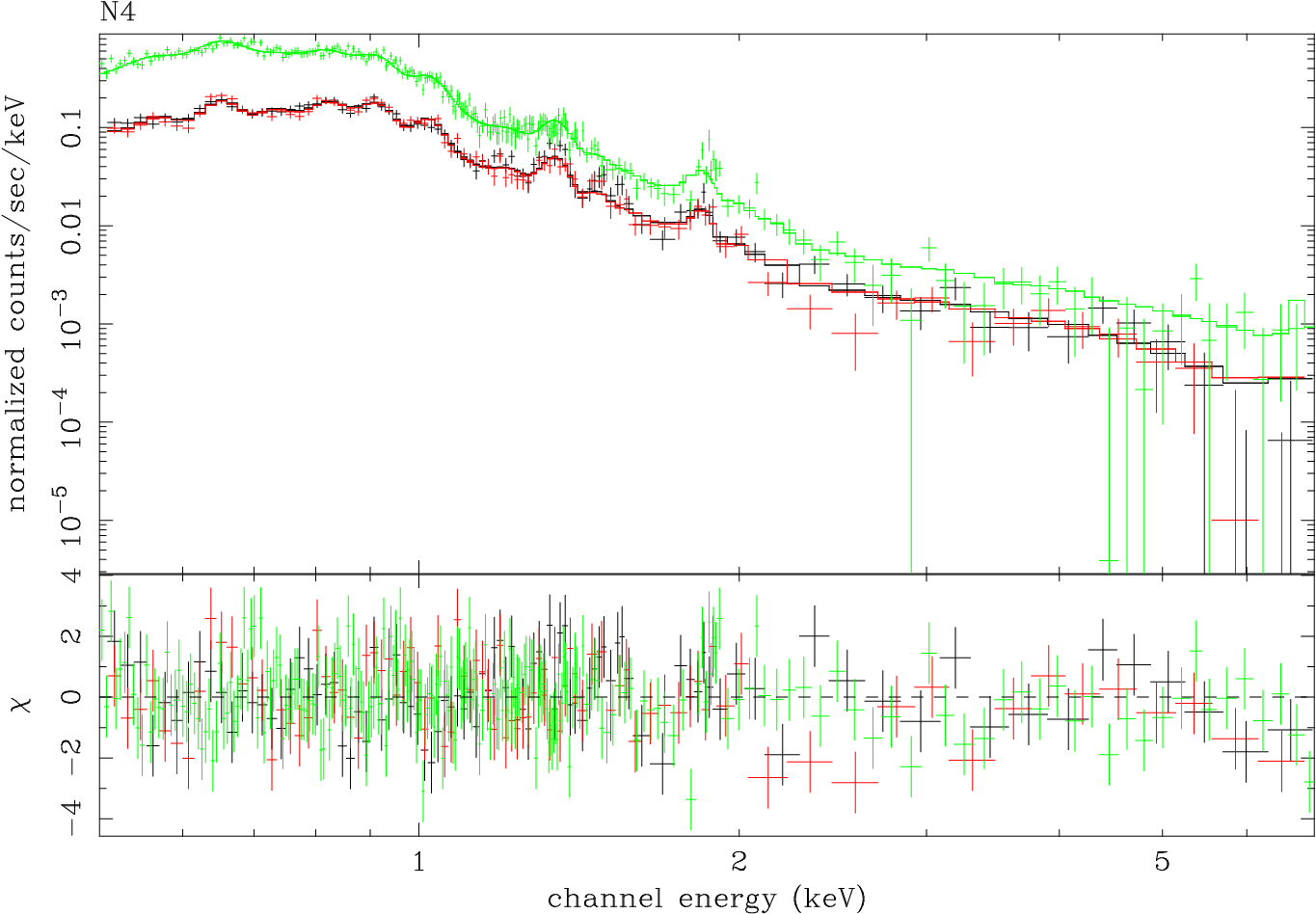}\\
  \includegraphics[width=.25\textwidth,angle=-90]{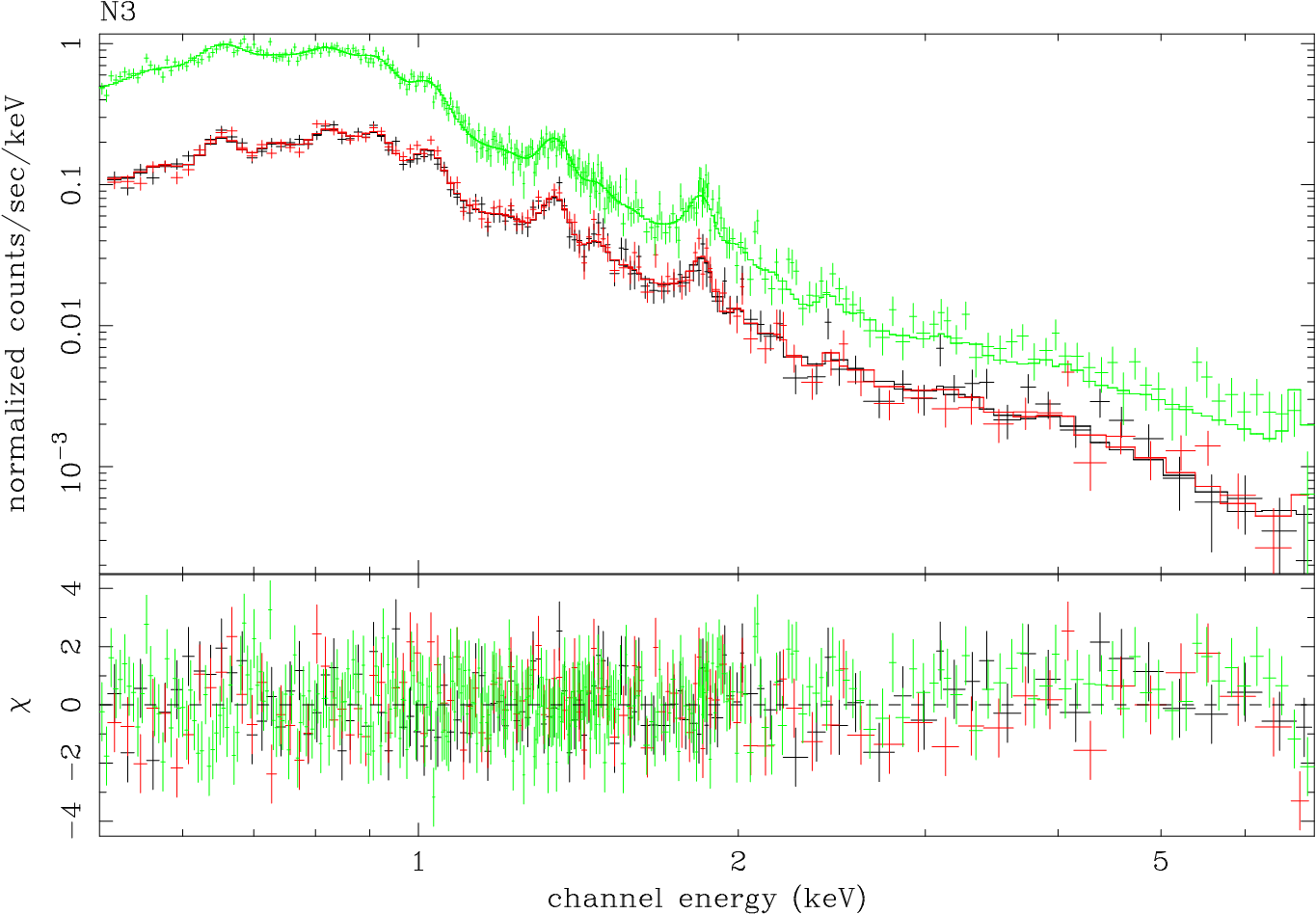}
  \includegraphics[width=.25\textwidth,angle=-90]{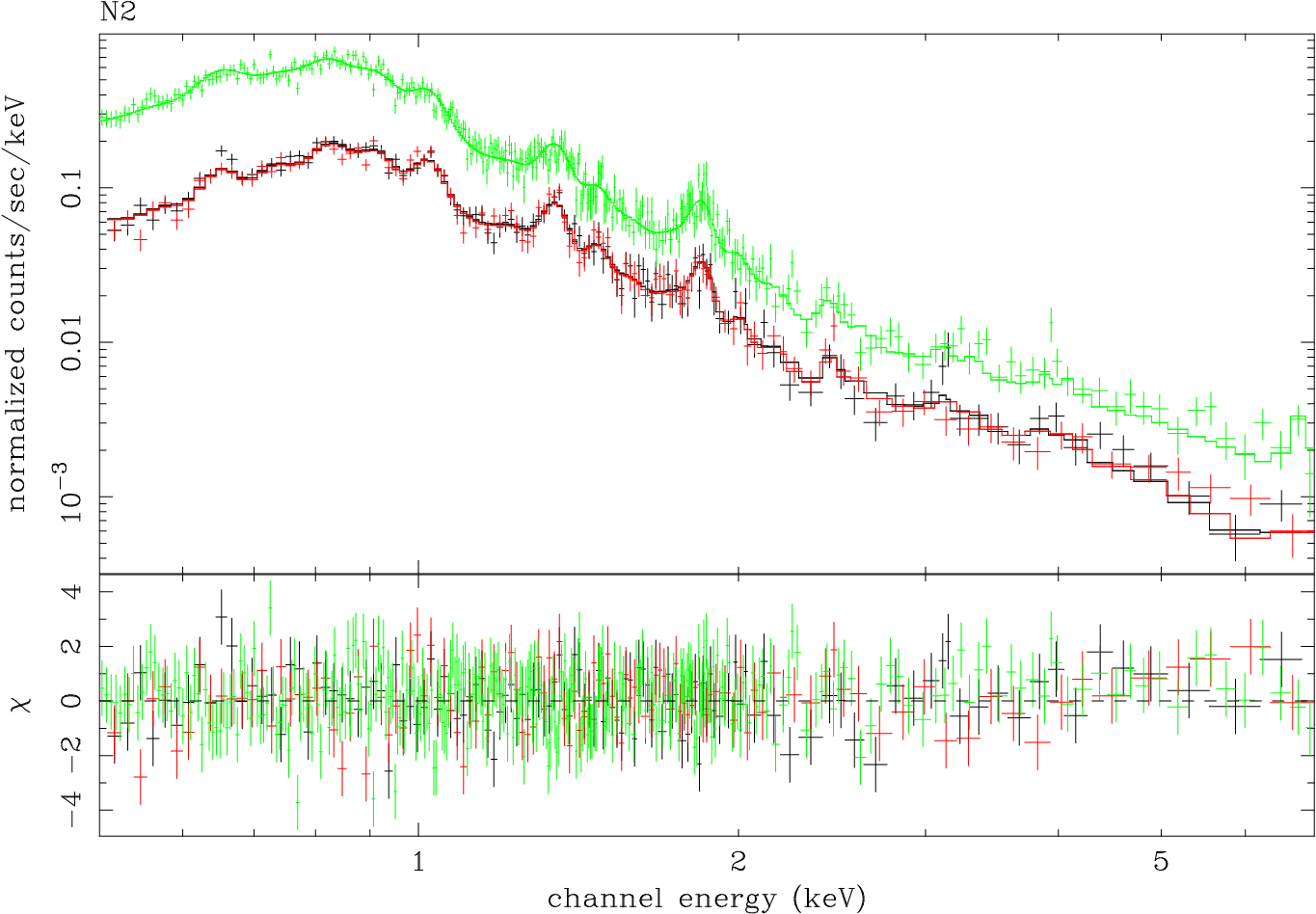}\\
  \includegraphics[width=.25\textwidth,angle=-90]{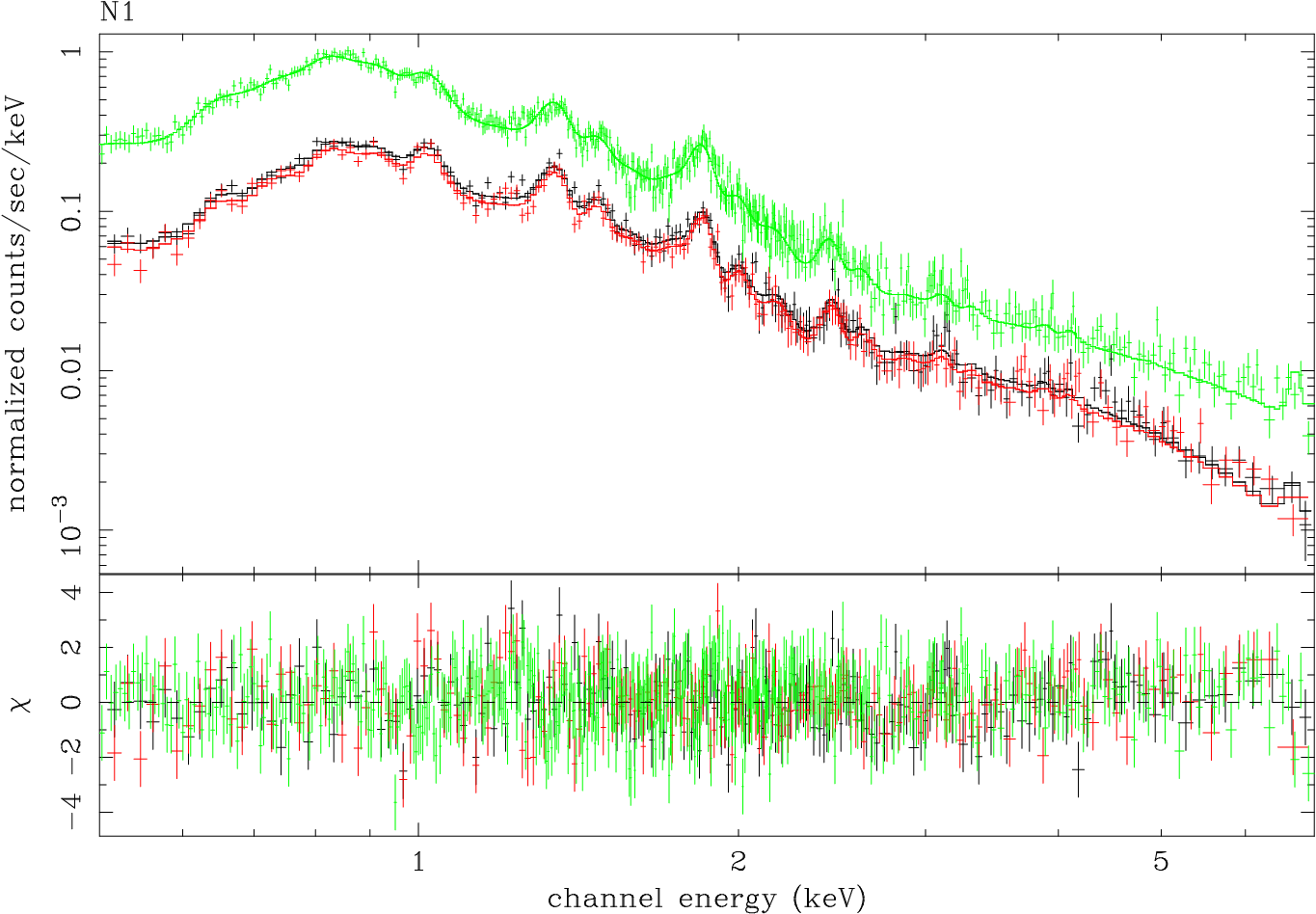}
  \includegraphics[width=.25\textwidth,angle=-90]{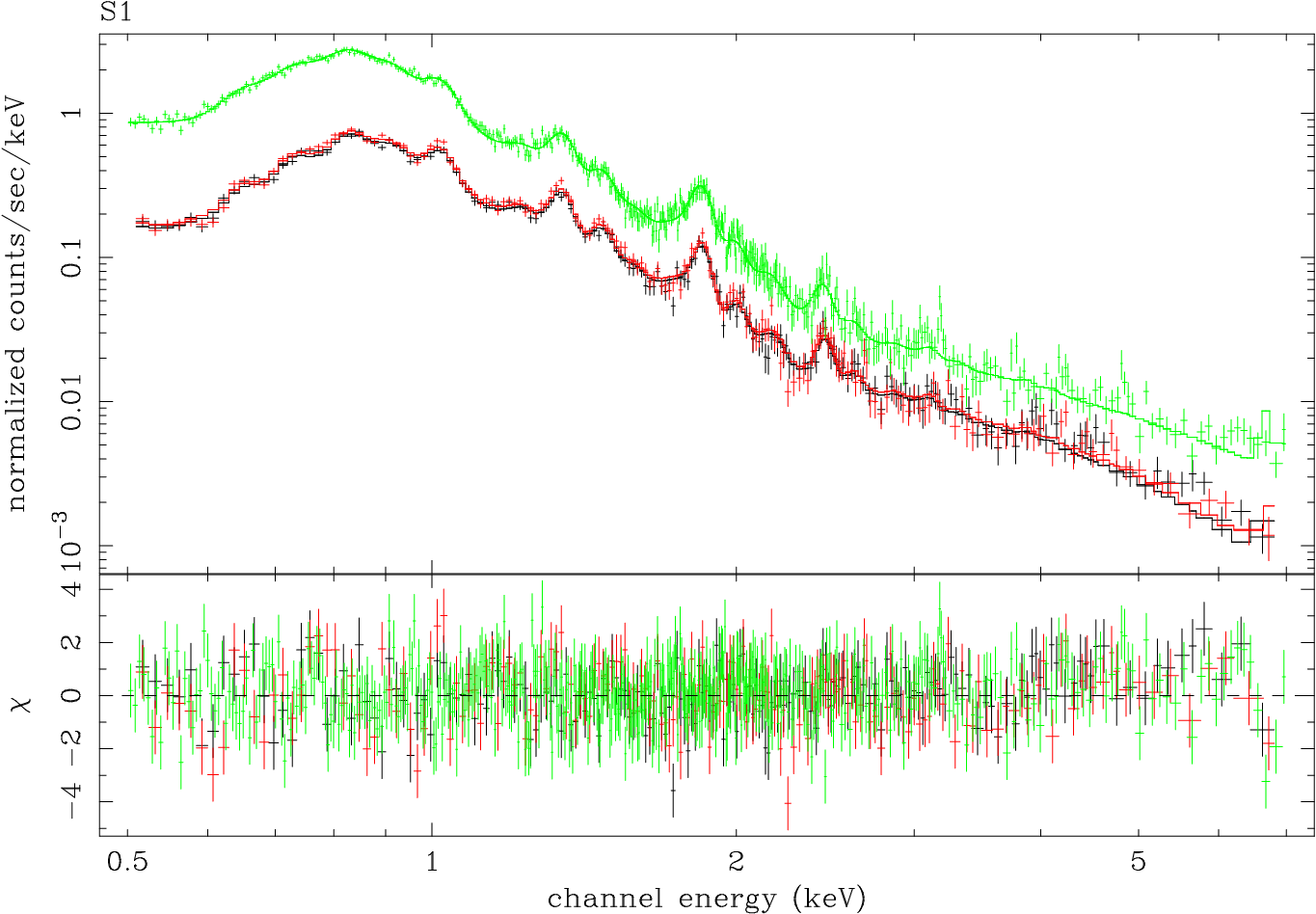}\\
  \includegraphics[width=.25\textwidth,angle=-90]{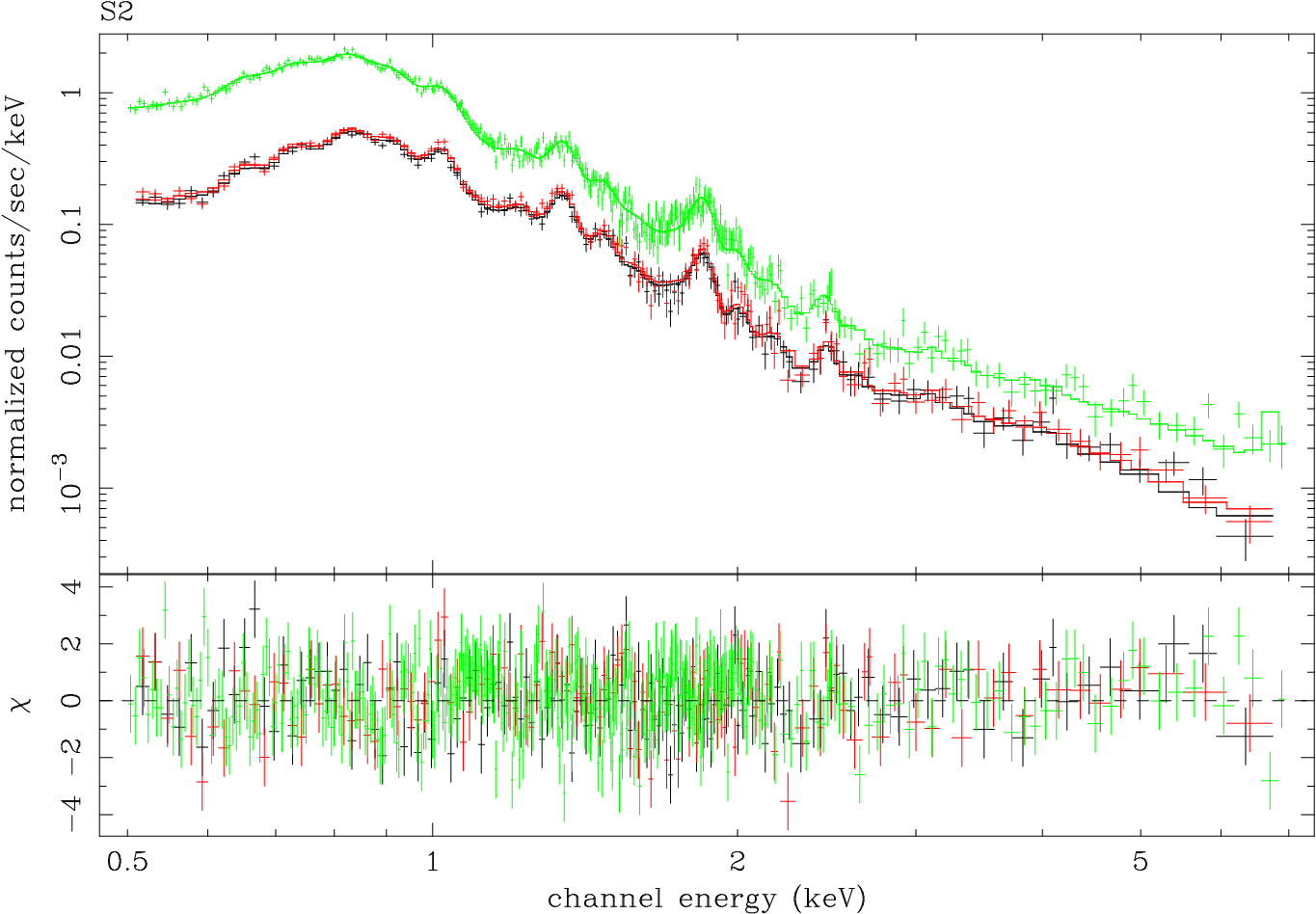}
  \includegraphics[width=.25\textwidth,angle=-90]{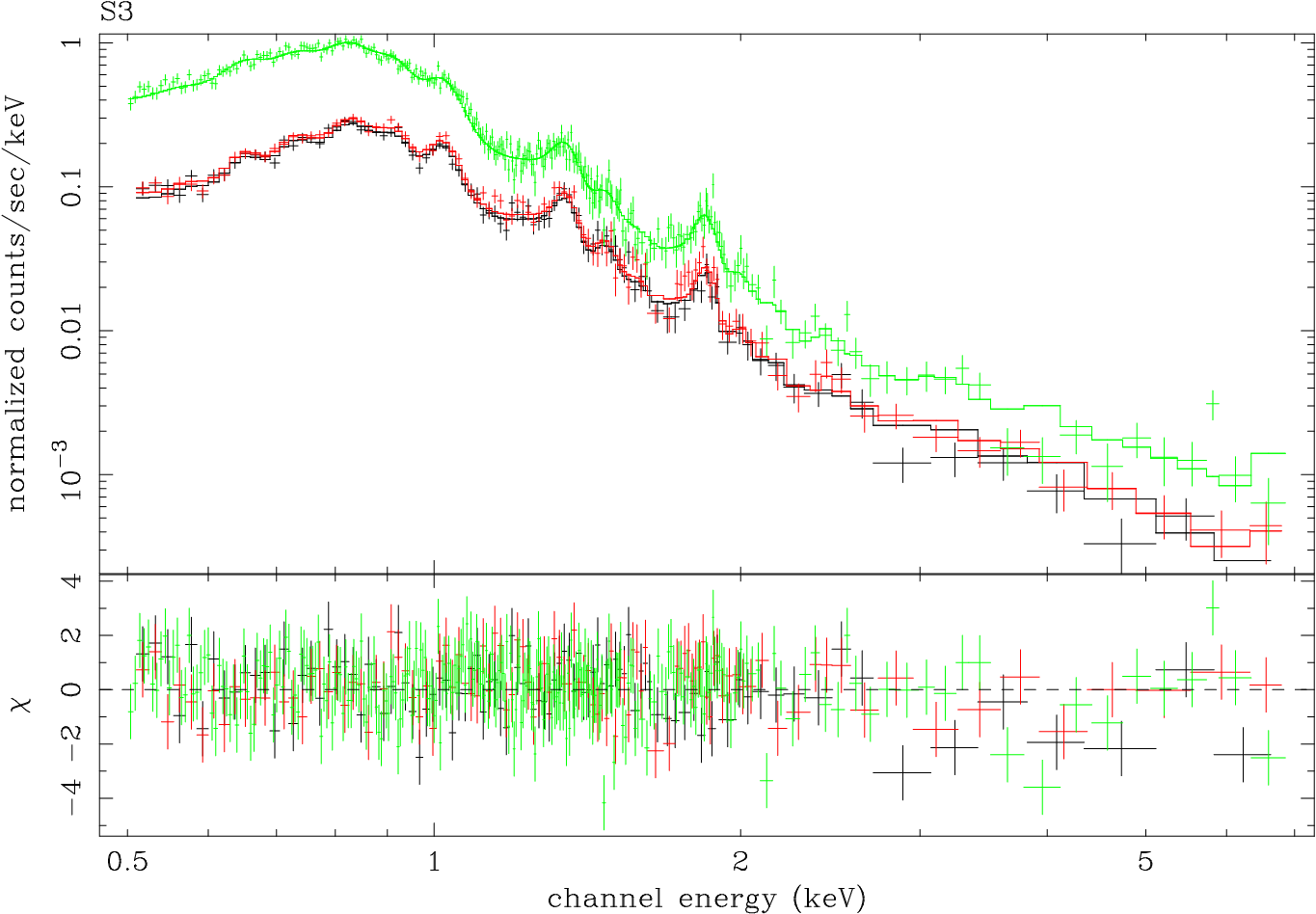}\\
  \includegraphics[width=.25\textwidth,angle=-90]{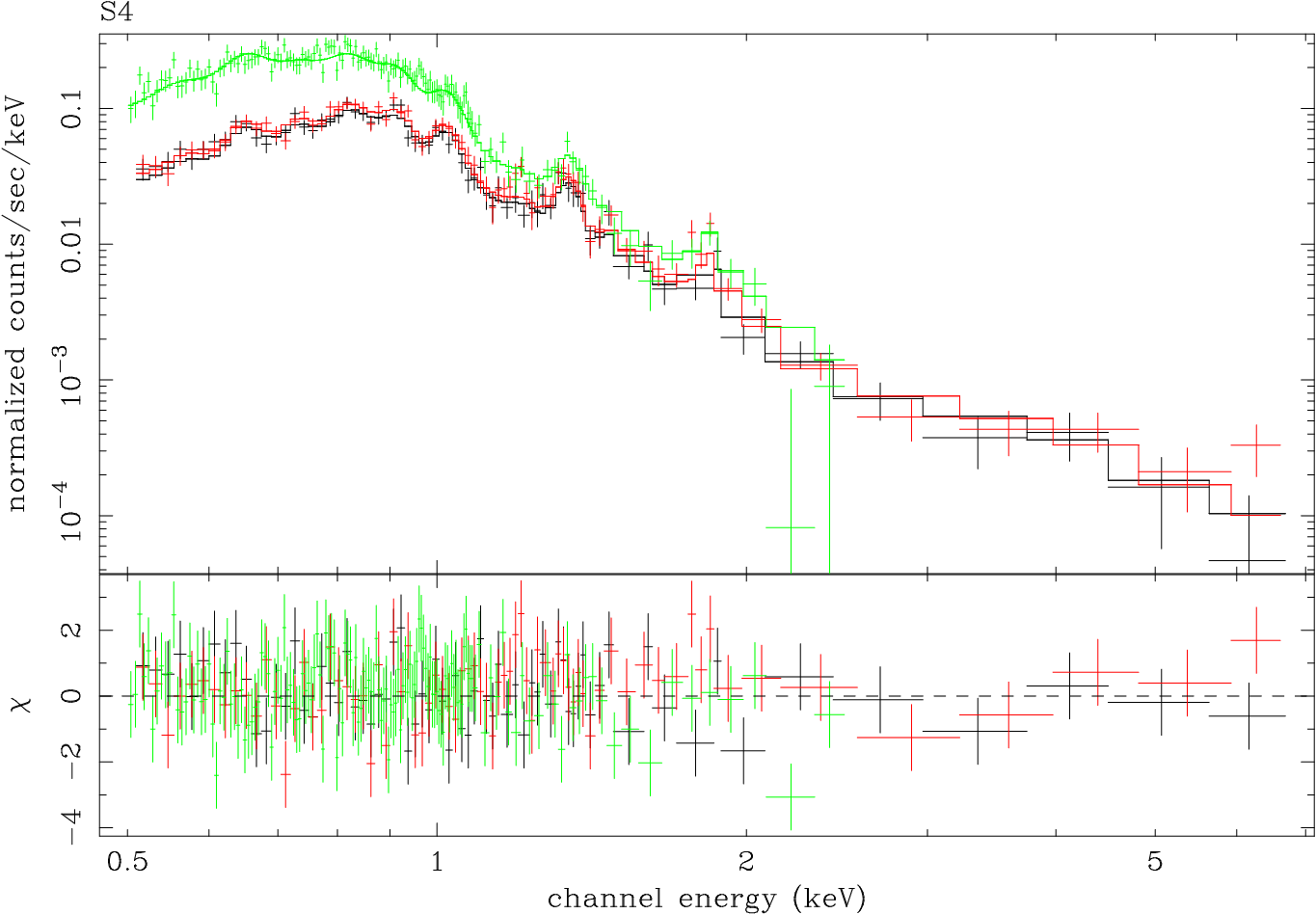}
  \includegraphics[width=.25\textwidth,angle=-90]{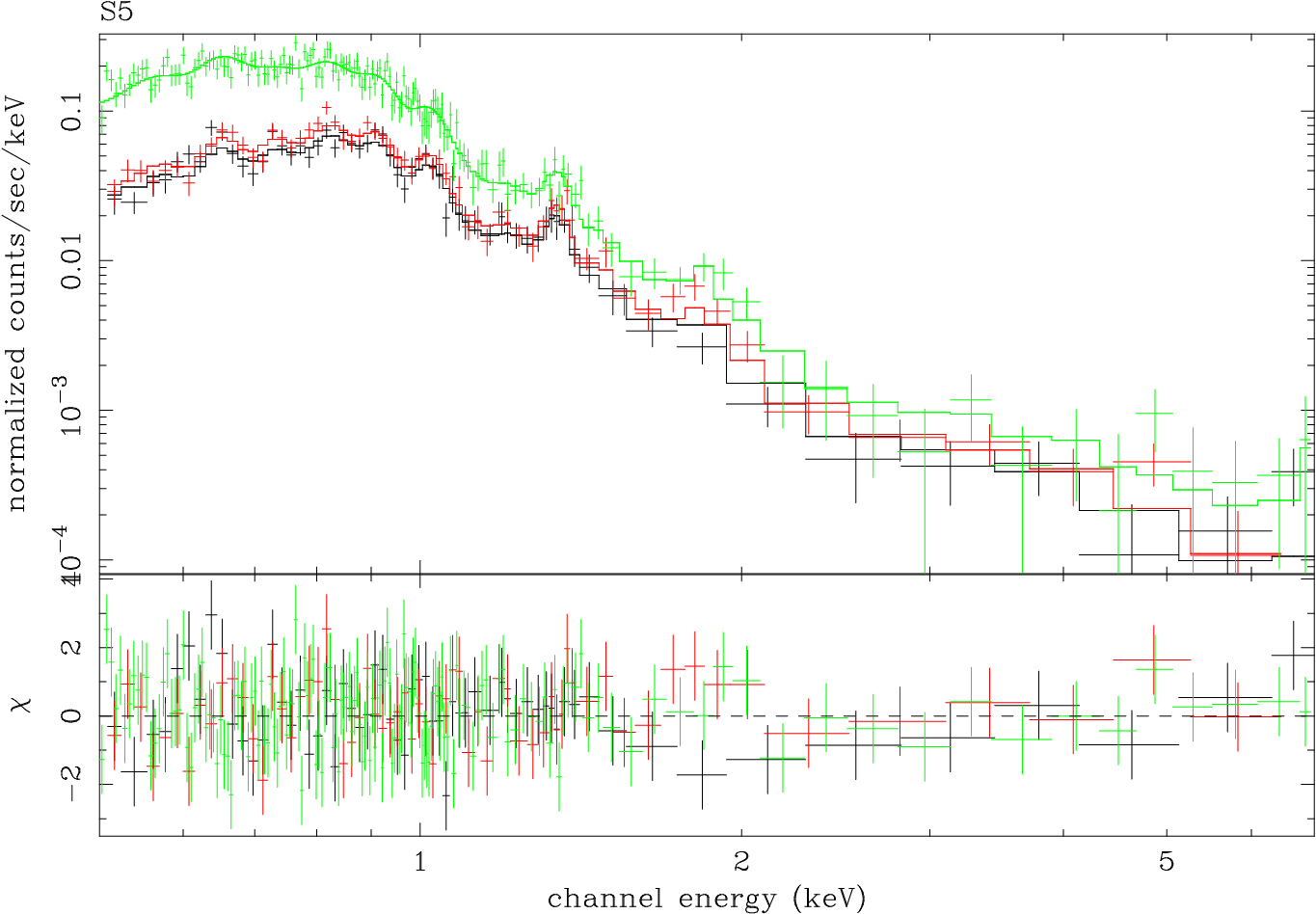}
  \caption{EPIC spectra for increasing height on
    the galactic plane. Green points (upper data series): {\em pn}. Black
    and red points (lower data series): MOS1 and MOS2, respectively.}
  \label{fig:atlantespettri}
\end{figure*}

To study the different properties of the hot gas as it flows and is
heated from the central starburst towards the intergalactic space, we
divided both outflows in five regions each. Each region has a
rectangular shape, the larger side being parallel to the galaxy major
axis.  The regions are numbered from N1 to N5, and from S1 to S5 with
increasing height above the galactic plane as shown in
Fig.~\ref{fig:xmmregioni}; in the figures we adopt the convention of
assigning negative heights to the northern regions. The regions have
different sizes, chosen mainly to guarantee enough photon counts and a
sufficiently good signal to noise ratio to perform a meaningful spectral
analysis in each area.  The spectra were extracted from the MOS and {\em pn}
data, and
fitted with the same models used for the central region, the only
difference being point sources, which in the outflow are sufficiently
apart to be excluded when extracting spectra. The positions of point
sources were determined by running the PWXDETECT wavelet-based
detection code on the MOS1, MOS2 and {\em pn} images \citep[the algorithm is
described in][]{damiani97}.  An atlas of the spectra with best-fitting
models is shown in Fig.~\ref{fig:atlantespettri}.  The APEC model was
used; no significant difference was found using the Mekal one.

\begin{figure}
  \centering
  \includegraphics[width=\columnwidth]{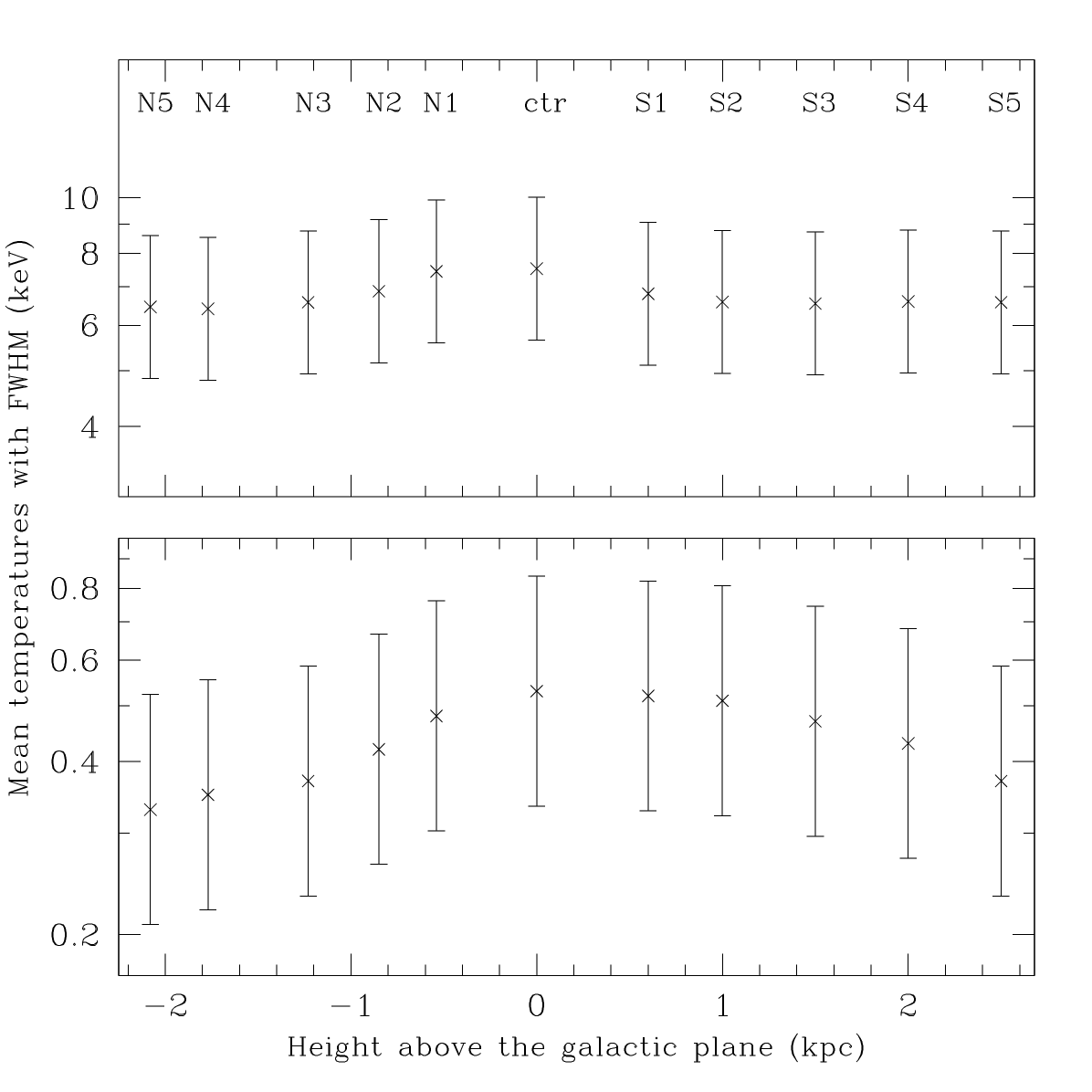}  
  \caption{Variation of temperatures of the plasma along the M82
    outflow. The lower and upper panels show the first ($\sim 0.5$
    keV) and second ($\sim 7$ keV) peak of the DEM, respectively.  The
    error bars represent the FWHM of Gaussian curves fitted to the
    peaks of the polynomial distribution of the DEM. For instance, the
    two error bars at 0 kpc and labelled `ctr' (i.e.\ {\em centre})
    are just a different representation of the right panel of
    Fig.~\ref{fig:DEM}. Negative values of distance refer to the north
    direction, positive values to south.}
  \label{fig:dem_tutte}
\end{figure}

The same multi-temperature structure is found in every region.
By fitting the DEM polynomial with two Gaussian components, we found
that the centre of the colder component moves slightly from 0.53 keV
in the centre of M82 to 0.33 and 0.37 in the outer regions (N5 and S5,
respectively). The hotter component does not significantly change its
temperature. Fig.~\ref{fig:dem_tutte} shows the centres and FWHM of the
best-fitting Gaussians.

\begin{figure*}
  \centering
  \includegraphics[width=.32\textwidth]{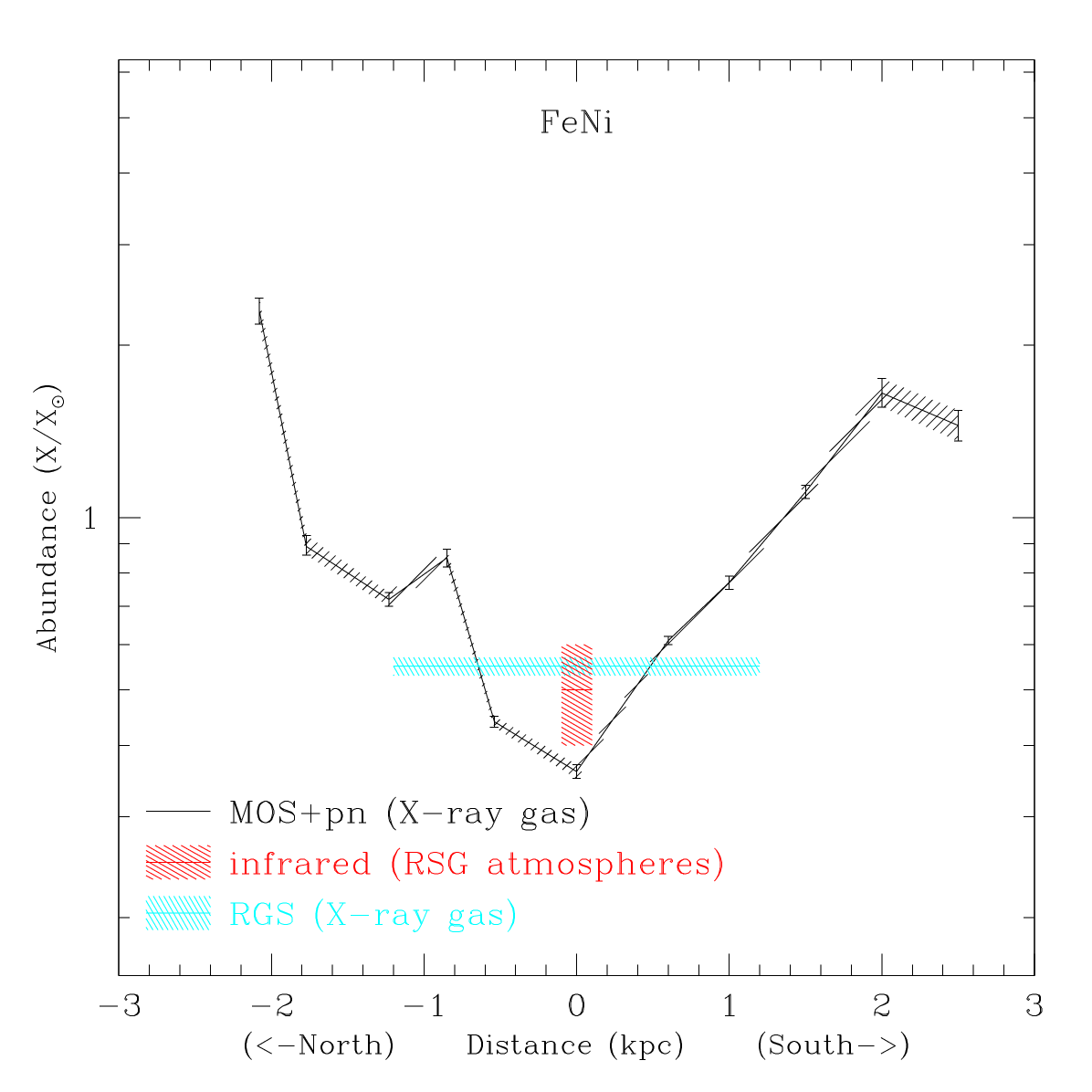}
  \includegraphics[width=.32\textwidth]{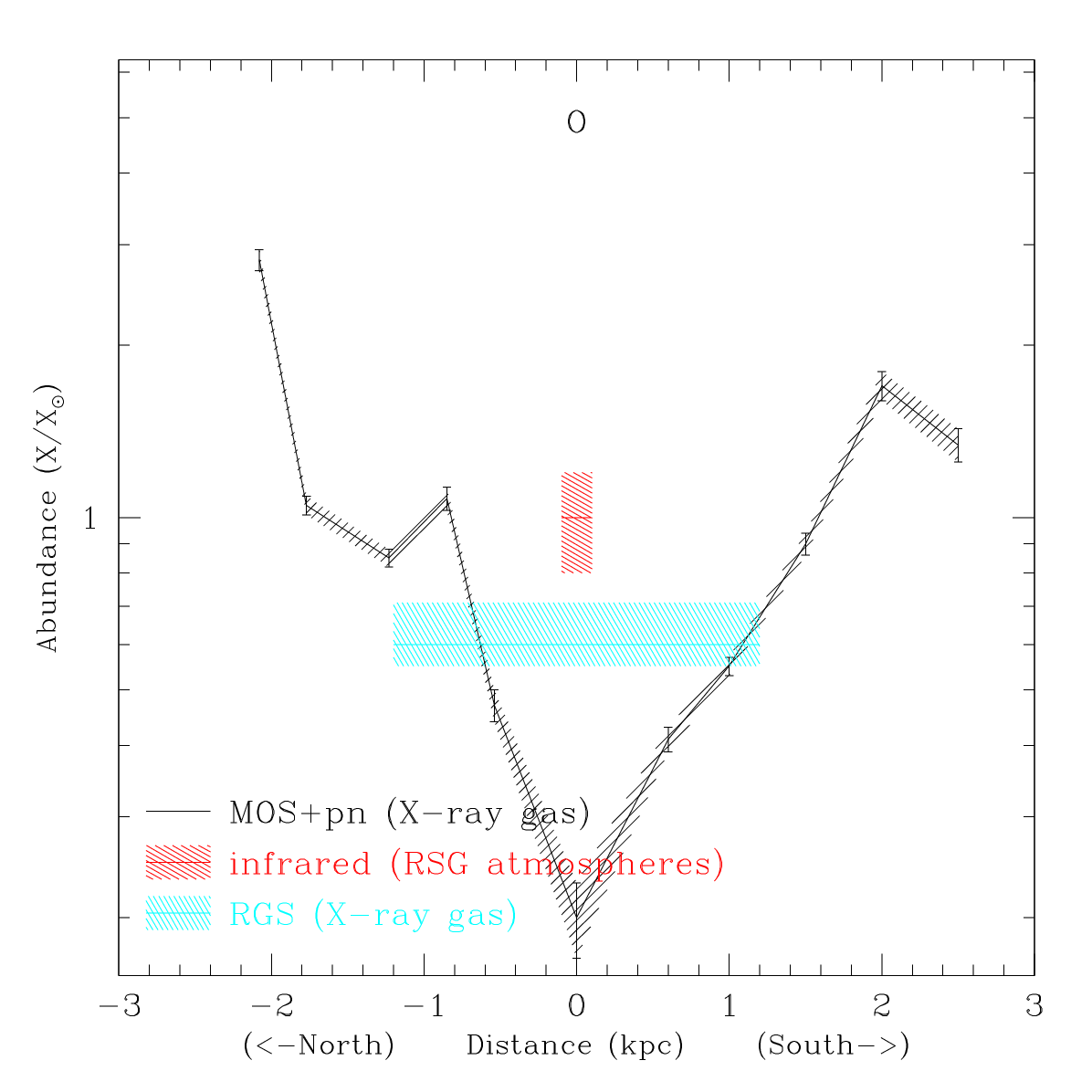}\\
  \includegraphics[width=.32\textwidth]{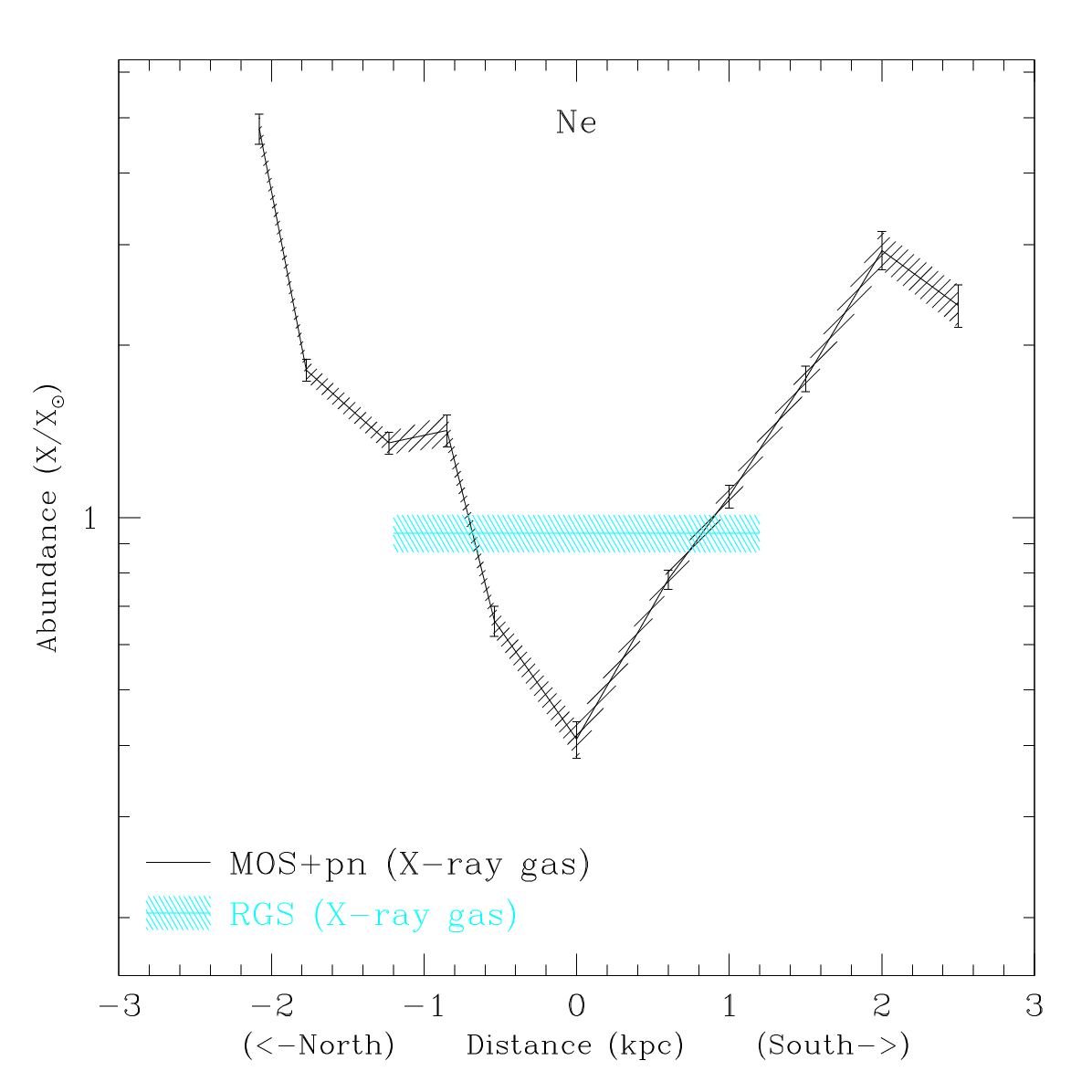}
  \includegraphics[width=.32\textwidth]{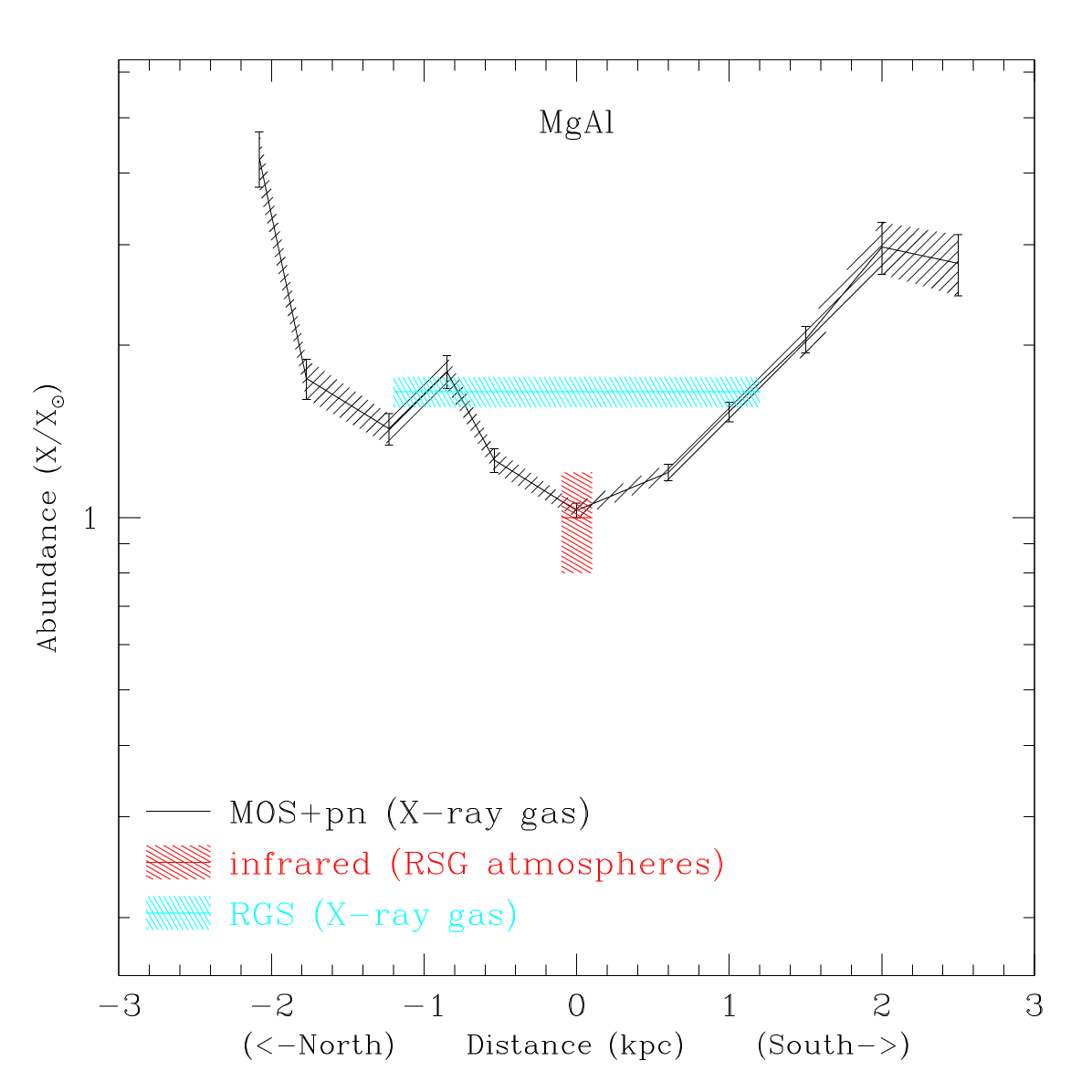}\\
  \includegraphics[width=.32\textwidth]{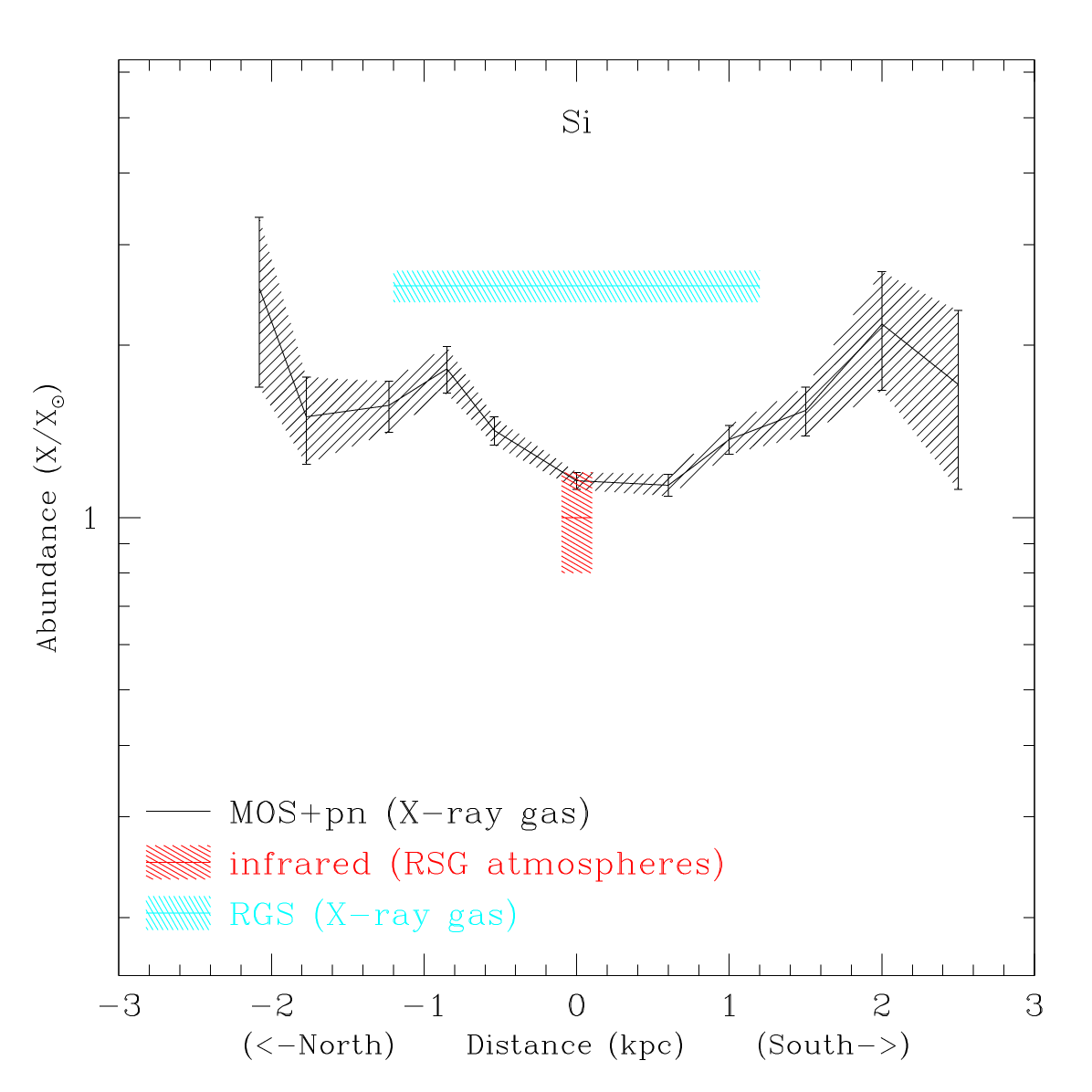}
  \includegraphics[width=.32\textwidth]{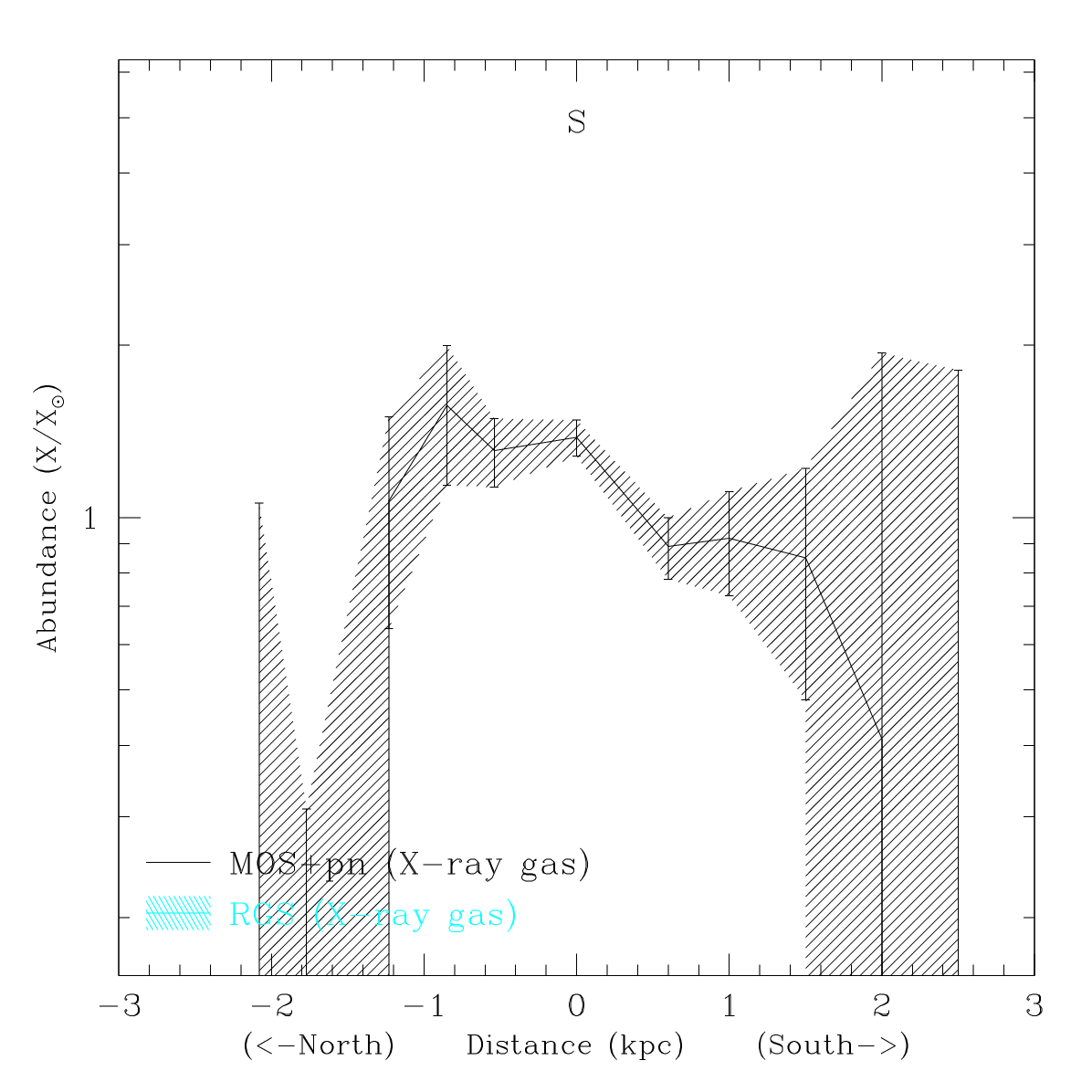}
  \caption{Variation of chemical abundances with increasing height on
    the galactic plane. Black: abundances from X-ray MOS and {\em pn}
    data. Blue: abundances from X-ray RGS data (due to the
    characteristics of the RGS spectrometer, they represent
    space-averaged values). Red: abundances from infrared data
    (corresponding to red supergiant stars in the galaxy central
    region). Negative values of distance refer to the north direction,
    positive values to south. }
  \label{fig:logelem}
\end{figure*}

The best-fitting chemical abundances from EPIC and RGS data are shown in
Fig.~\ref{fig:logelem} along with results from infrared observations
for the central regions \citep{or04}.  Only one region (S1) required
in its spectrum some absorption in excess of the Galactic value,
probably due to the M82 disc partially covering the southern outflow.
We confirm our previous finding \citep{or04}, that the lighter \ael\ 
are more concentrated in the outflow than in the centre. This effect
is larger for elements with lower atomic mass, becomes less
evident for Si and reverses for S.  The centre/outskirt abundance
ratio in the centre is about $\sim 1/10$ for O and Ne. Fe is also more
concentrated in the outflow.

\begin{figure}
  \centering
  \includegraphics[width=\columnwidth,bb=32 148 400 698,clip]{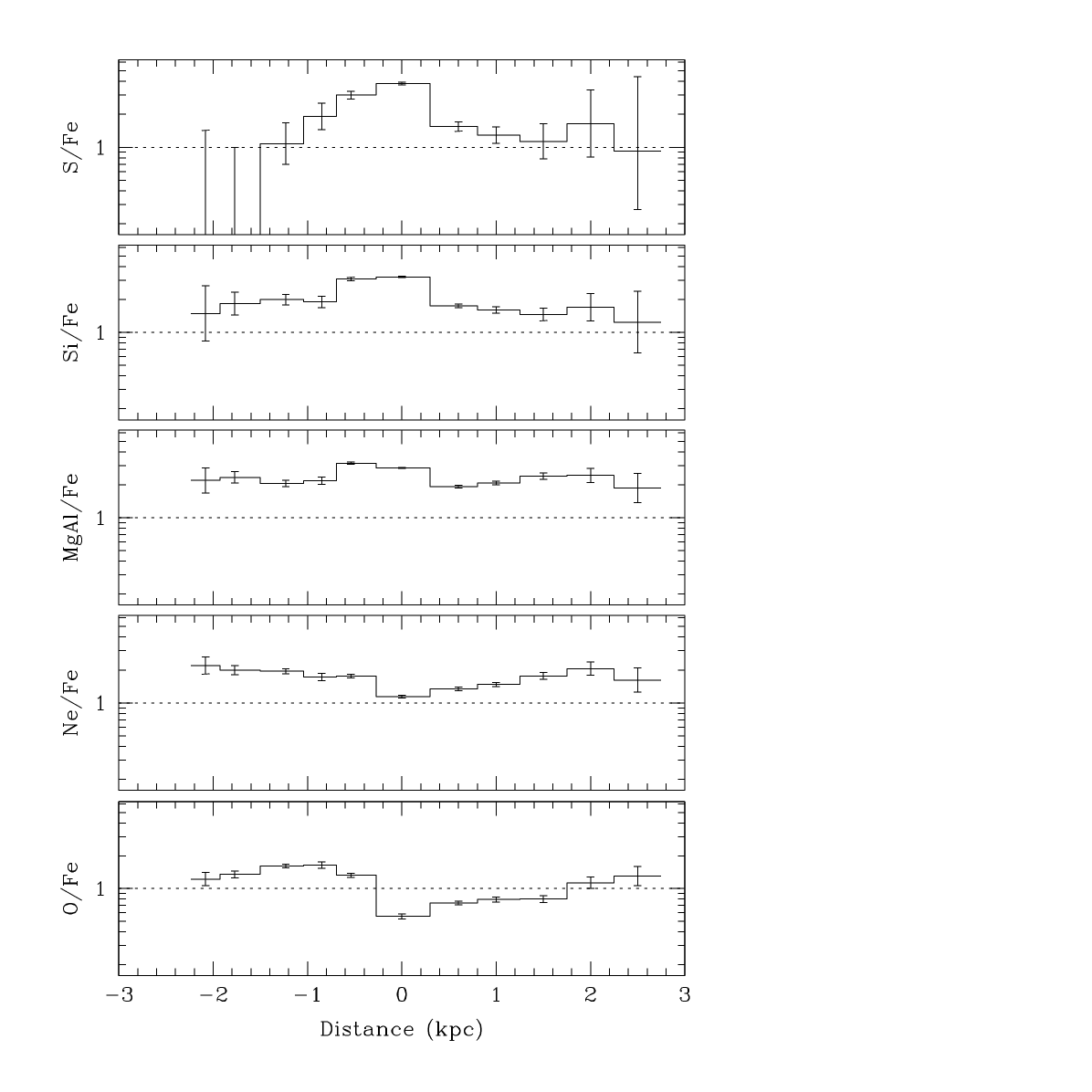}  
  \caption{Abundance ratios (X/Fe) observed in EPIC spectra of the
    outflow. Negative values of height refer to the north
    direction, positive values to south.}
  \label{fig:xfe}
\end{figure}

The abundances ratios, shown in Fig.~\ref{fig:xfe}, have smaller
variations, and present different trends for light and massive
elements: while the O/Fe and Ne/Fe ratios are lower in the centre than
in the outskirts, the opposite holds for Si/Fe and S/Fe, with Mg/Fe
being an intermediate case showing no variation. The scatter between
values for centre and outskirts is a factor of $\lesssim 3$.

The physical parameters of the plasma may be obtained from the
temperature ($T$) and normalisation ($A$) of the model, with some
assumptions about the volume ($V$) and filling factor ($f$). The
normalisation of plasma models in XSPEC is defined as 
\begin{equation}
\label{eq:unitanorm}
A=\frac{ 10^{-14} }{
4\pi (D_{\rm A} (1+z))^{-2}\cdot \int n_{\rm e} n_{\rm H} \de V }
\end{equation}
using c.g.s.\ units and where $D_{\rm A}$ is the angular distance, and
$n_{\rm e}$ and $n_{\rm H}$ are the electron and hydrogen volume
densities.  One usually assumes $n_{\rm e}\simeq n_{\rm H}$, and
defines the Emission Integral (EI) as $EI=\int n^2 \de V$.  From this,
one can obtain the density $n_{\rm e}\simeq (EI / V f)^{1/2}$, the
pressure $p\simeq 2 n_{\rm e} kT$, the gaseous mass $M\simeq n_{\rm e}
m_{\rm p} V f$, the thermal energy content $E\simeq 3 n_{\rm e} kT
Vf$, and the radiative cooling time $t\simeq 3 kT / (\Lambda n_{\rm
  e})$, where $m_{\rm p}$ is the proton mass, and $\Lambda$ is the
emissivity of the gas. Since $\Lambda$ depends on the temperature, its
values were interpolated from Table~6 of \citet{sutherland93}
($\Lambda$ is of the order of $\sim 3\e{-23}$ erg cm$^3$ s$^{-1}$,
valid for solar metallicity and the temperatures of interest).

The parameters were calculated using the best-fitting normalisations for
each region. The used temperatures are the best-fittings to the low-energy
components of the DEMs (see lower panel in Fig.~\ref{fig:dem_tutte}).
For each of the rectangular N and S regions
(Fig.~\ref{fig:xmmregioni}), the volume was assumed to be that of a
cylinder with base diameter and height equal to the major and minor
sides of the rectangle, respectively.  Normalisations, volumes and
physical parameters of the plasma are shown in Table~\ref{tab:fisica}.
The values for the galaxy centre are also shown in the same table,
assuming a cylindrical volume with base radius $32\arcsec$ and height
$120\arcsec$. From the Table, one may see that the gas density and
pressure decrease by a factor of $\sim 10$ from the centre to the
outskirts, while the cooling time increases.

\begin{table*}
  \centering
  \begin{tabular}{lrrrrrrrrrrr}
\hline  
                       &N5     &N4    &N3   &N2   &N1    &centre &S1   &S2    &S3    &S4    &S5   \\
\hline
  Normalisation ($10^{-5}$)
                       &0.14   &0.68  &1.6  &1.6  &4.0   &27     &5.3  &2.4   &0.77  &0.14  &0.18 \\
  Volume (kpc$^3$)     &0.61   &1.2   &1.1  &0.43 &0.41  &1.4    &0.24 &0.35  &0.26  &0.21  &0.22
                                                                                         \smallskip \\
  Density ($10^{-3}$ cm$^{-3}$)
                       &$3.4 f^{-\frac{1}{2}}$
                               &5.4   &8.7  &14   &23    &32     &35   &19    &13    &6.0   &6.5  \\
  Pressure ($10^{-12}$ dine cm$^{-2}$)
                       &$3.6 f^{-\frac{1}{2}}$
                               &6.1   &10   &19   &35    &54     &58   &31    &19    &8.3   &7.7  \\
  Mass ($10^5$ M$_\odot$)
                       &$0.52  f^{+\frac{1}{2}}$
                               &1.7   &2.4  &1.5  &2.3   &11     &2.0  &1.7   &0.81  &0.31  &0.37 \\
  Energy ($10^{53}$ erg)
                       &$0.99  f^{+\frac{1}{2}}$
                               &3.3   &5.2  &3.6  &6.3   &33     &6.0  &4.8   &2.2   &0.76  &0.77 \\
  Energy density ($10^{-12}$ erg cm$^{-3}$)
                       &$5.5   f^{+\frac{1}{2}}$
                               &9.2   &15   &28   &52    &81     &87   &47    &28    &12    &12   \\
  Cooling time (Myr)
                       &$450   f^{+\frac{1}{2}}$
                               &310   &210  &160  &120   &100    &89   &160   &220   &390   &280  \\
\hline
  \end{tabular}
  \caption{Physical parameters of the plasma across the different
    regions of the outflow. The dependencies on the filling factor $f$
    have been explicited for simplicity only in the first column, but
    they apply to all columns.}
  \label{tab:fisica}
\end{table*}

\section{Discussion}
\label{sec:discussione}

\subsection{Multi-temperature nature of the gas}
\label{sec:disc:multitemp}
To find why the DEM has a double-peaked shape (Figs.~\ref{fig:DEM} and
\ref{fig:dem_tutte}), we first checked if the hot component (the peak
at $\sim 5$ keV) was really required to achieve a good fit. By
truncating the DEM and leaving only the low energy ($\sim 0.5$ keV)
peak, no similar or better fit than those shown in
Table~\ref{tab:src0} could be obtained, ($\chiq\sim 3028$ with 2273
degrees of freedom, $\chiqr\sim 1.33$, CE lines included).  Then we
checked if the hot component could have a non-thermal origin. Since
star forming galaxies are known as radio sources, bremsstrahlung
emission should in principle arise from the same population of
electrons from which the synchrotron emission is originated. Thus we
developed a non-thermal bremsstrahlung model for XSPEC following the
prescriptions in \citet{blumenthalgould}.  By including such a
component in the model, no improvement of the fit statistics was
observed with respect to the previous case (i.e., no hot component at
all).

It is unlikely that the hot component is due to unresolved
point sources. By analysing the \chandra\ 3--9 keV spectra of the
diffuse emission of M82, and by considering the \lognlogs\ of point
sources in star forming galaxy, \citet{strickland07} have shown that
significant flux remains in that band after accounting for unresolved
sources.

Thus, assuming that both components were thermal, one might naively
identify the two temperatures with plasma coming from different
regions, but this idea would have a hard time explaining why the
temperatures are about the same across all the outflow, and also why a
similar pattern with same temperatures seems to be exhibited by other
galaxies, e.g.\ NGC253.

The temperature pattern exhibited by M82 is strikingly similar to that
of the Galactic ridge (GR), which is an enhancement of the X-ray
background along the Galactic equator \citep{cooke69}, and whose
explanations range from truly diffuse emission to the sum of many
faint sources.  Unlike the GR case, an interpretation based on the
hypothesis of unresolved point sources does not seem viable for M82.
The main evidence against is the morphology of the outflow, which is
not associated with any stellar structure, such as the disc or the
halo.

The temperature pattern is not the only similarity between M82 and the
GR: another, important one is the energy of the Fe K lines.  By using
a simple model (power-law plus Gaussian line) for the 4-8 keV EPIC
spectrum, an Fe thermal line was detected%
\footnote{This line is fully reproduced by the MEKAL and APEC models
of Sect.~\ref{s:spettri1}.} 
at $6.66\pm 0.02$ keV ($\sigma\le 0.08$, $\chiqr\sim 622/568\sim
1.09$) with equivalent width $\sim 100$ eV (thus confirming the value
originally found with BeppoSAX by \citealt{cappi99}). A similar value is
found in \suzaku\ data (P. Ranalli, in prep.). 

For the GR, \citet{koyama86} found a line energy of $6.71\pm 0.04$ keV
in observations with the {\em Tenma} satellite. \citet{kaneda97}
reported a line energy of $\sim 6.6$ keV.  The best-fitting line width
also seem to agree (GR: $\sigma\sim 70$ eV; M82: $\sigma\lesssim 80
eV$).  One should also note, however, that \citet{tanaka02} reported a
higher energy for the Fe line in the GR ($6.76$ keV), and that a line
centroid of $\sim 6.66$ keV is also consistent with a CE origin for
the Fe XXV triplet \citep{wargelin05}.

Anyway, the presence of hot gas with $kT\sim 7$ keV should put a
significant number of Fe atoms in the highest ionisation states,
namely He- and H-like, whose line emission happens at higher energies
($\sim 6.8$ keV). Since the presence of such a line in M82 can be
safely excluded, one might ask if the hot plasma component is indeed
real or is just an artefact.

An important insight has been provided by \citet{dogielmasai02} and
\citet{masaidogiel02}. The authors argue that the bremsstrahlung
emission of electrons from the regions in which particle acceleration
occurs has a peculiar spectrum which mimics the double-temperature
pattern found both in the GR and in M82. By considering {\em in situ}
acceleration of thermalised electrons, they derive a broad transition
region between the thermal and non-thermal energy regions.  The total
X-ray spectrum resulting from both the thermal and transition regions
of the electron energy spectrum may then be described, with good
approximation, by the sum of two thermal plasmas whose temperatures
are in the ratio $T_{\rm hotter}/T_{\rm colder}\sim $4--5. In this
model, the actual temperature of the thermalised gas is the colder of
the two (the authors propose a value of 0.3 keV for the GR).  This
would then naturally explain the measured energy of the Fe K line.
Another consequence of this model would be the presence of synchrotron
emission from the outflow, which in M82 is already known
\citep{seaquist91}.  However, the $T_{\rm hotter}/T_{\rm colder}$
ratios observed in M82 are larger than the predicted one by a factor
3--4 (see Fig.~\ref{fig:dem_tutte}).  While the model looks promising,
some more work is still needed to check if the discrepancy in the
temperature ratio can be reconciled.

\subsection{Vertical abundance gradients}
\label{sec:disc:grad-abbond}
In a somewhat naive scheme, a cavity of hot enriched gas is
created in the ISM when the first supernova explosions begin. The hot
gas which happens to be at the boundary between the cavity and the
cold ISM may either cool and mix, or stay hot and unmixed. Let us
suppose it does not mix significantly until the hot gas bubble breaks
out of the galaxy plane and the outflow is formed. Since the more
massive a SN progenitors is, the more metals are produced, the outer
regions of the outflow should have larger abundances than the inner
one.

\begin{figure}
  \centering
  \includegraphics[width=.8\columnwidth,bb=19 145 383 716,clip]{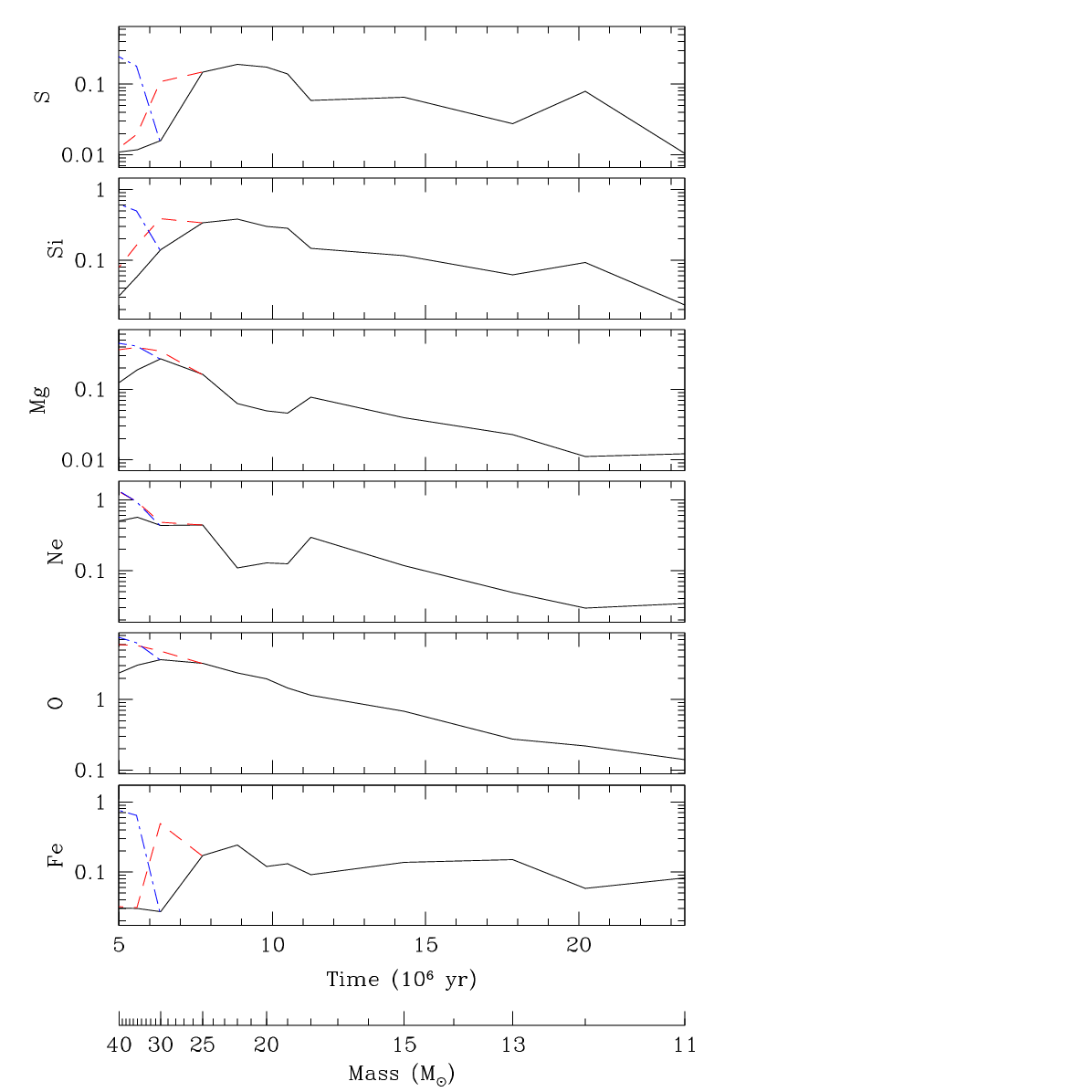}
  \caption{Masses (in $M_\odot$) of ejected material from type-II SN
    explosions of massive stars with solar metallicity, as functions
    of the lifetime in the main sequence (data from
    \citealt{ww95}). The (black) solid line refers to calculations
    made using the standard value of $1.e\e{51}$ erg for the final
    kinetic energy of the ejecta (model A in WW95). The (red) dashed
    and (blue) dot-dashed lines refer to more energetic explosions
    (models B and C in WW95) which affect only the most massive
    progenitors.  For clarity, a mass scale is also displayed
    \citep{traccedipadova}.}
  \label{fig:pseudorecchi}
\end{figure}

\begin{figure}
  \centering
  \includegraphics[width=.8\columnwidth,bb=19 145 383 716,clip]{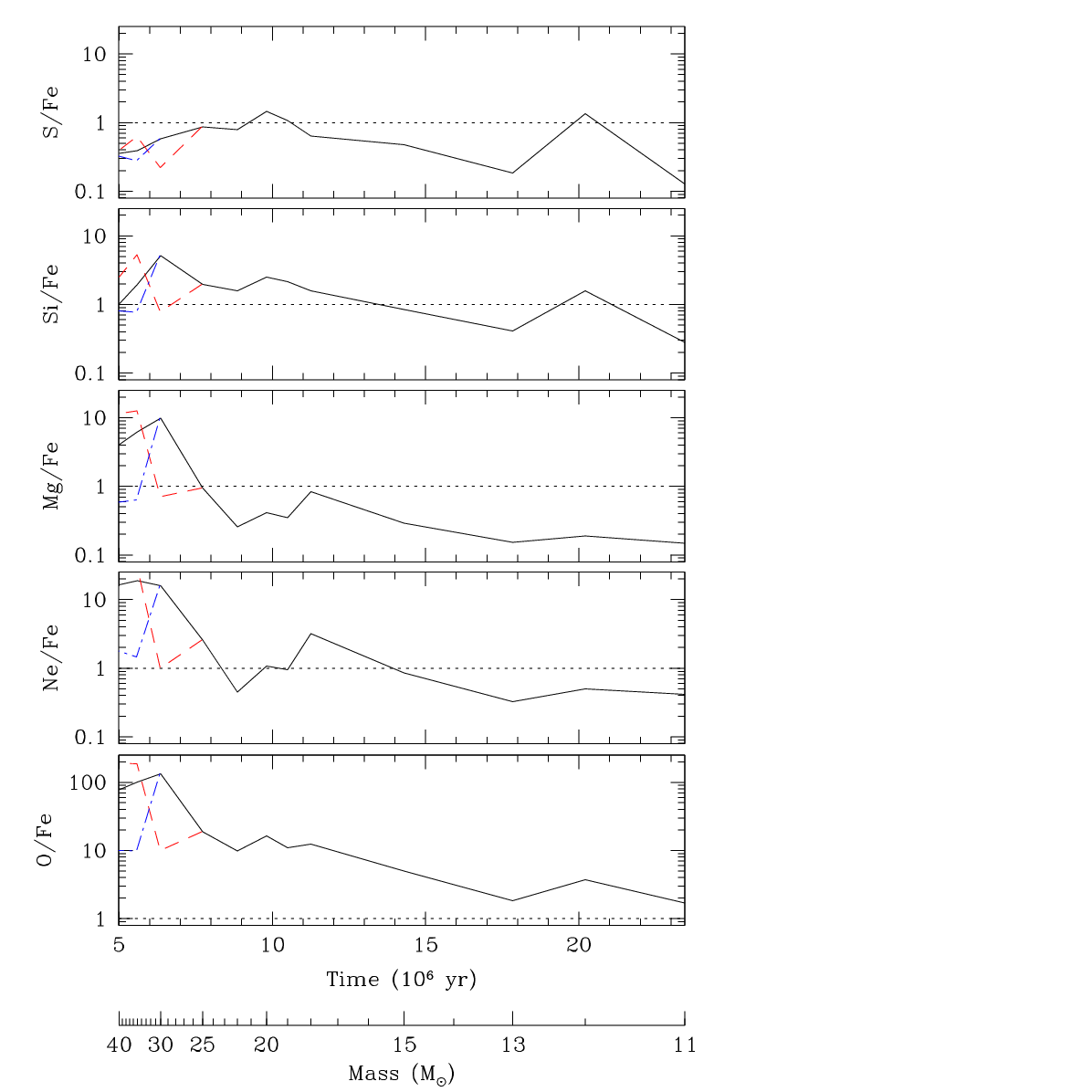}
  \caption{Abundance ratios ($M_{\rm X}/M_{\rm Y}$) of material
    ejected by type-II SN explosions of massive stars with solar
    metallicity, as functions of the lifetime in the main sequence
    (data from \citealt{ww95}). Colours, line styles and scales as in
    Fig.~\ref{fig:pseudorecchi}.}
  \label{fig:xfe-ww95}
\end{figure}

To set this idea in a quantitative framework, the relationships
between the lifetime of the SN progenitor and the mass of material
released after the explosion are shown in Fig.~\ref{fig:pseudorecchi},
with data from \citet[][hereafter WW95]{ww95} and evolutionary tracks
from \citet{traccedipadova}.  Among the several models present in WW95
we have considered a solar value for metallicity, since this meets
most closely the IR and X-ray values.  In Fig.~\ref{fig:xfe-ww95} we
also show the X/Fe mass ratio from WW95.

To locate the proper time-scale, it is important to consider the age of
the starburst in M82. 
\citet{matsushita00} report from CO measurements the presence of a
molecular super-bubble of size $\sim 120\times 240$ pc in the
central regions of M82, expanding at $\sim 100$ km s$^{-1}$, which lead them
to derive an age of 1--2 Myr. However, this bubble seems to have not
yet fully broken, so it may represent a starburst episode successive
to the one(s) that produced the X-ray emitting outflow.
\citet{strickland97} report from ROSAT
observations of the X-ray halo that the age may be 10 million years.
Also, \citet{silva01} investigated the radio to X-ray spectral
energy distribution in M82 and found a value for the starburst age of
$\lesssim 25$ Myr. 
The models in \citet{stricklandstevens00} suggest that the time
required for the starburst in M82 to break out of the ISM is about 5
million years. 
Finally, the presence of red supergiant stars in the centre of M82
\citep{or04} puts a lower limit to the starburst age at $\sim 7$ Myr.

Looking at Fig.~\ref{fig:pseudorecchi} one may find a global decrease
of the production of O, Ne, Mg, Si, S for increasing progenitor
lifetime (or, equivalently, for decreasing mass). In other words, the
first stars to explode are also the ones with larger yields. The Fe
production, on the other hand, has a less pronounced decrease, with a
secondary maximum at $\sim 18$ Myr; it also has large uncertainties
for short-lived stars.  It may be tempting to look again at
Fig.~\ref{fig:logelem}, under the hypothesis that the emission further
away from the galaxy centre traces the ejecta from more massive and
short-lived stars. This picture may be checked also by comparing the
observed and theoretical abundance ratios (Figs.~\ref{fig:xfe} and
\ref{fig:xfe-ww95}).
However, several pitfalls may hamper this interpretation. For
instance, we have ignored here any role of the IMF or of the mixing;
this is only true insofar as one assumes that the outflow gas is produced
in an instantaneous starburst, and then moves in bulk with a constant
velocity and without running into other material, such as cold neutral
gas in the disc or hot halo gas. To cope with this difficulties would
require more assumptions about the history of the starburst-produced
gas and of its outflow than it is advisable to discuss here. For
instance, if the starburst is not instantaneous but rather has a
finite duration, then one needs to consider the IMF, whose effect is
to introduce a blurring in the yields distribution of
Fig.~\ref{fig:pseudorecchi}. Also, the mixing of the
starburst-produced gas with pre-existent halo gas might veer the
observed chemical abundances ratios from resembling type-II SN yields
towards type-Ia SN yields.

\medskip

Another possibility is that the M82 wind may be significantly more
mass-loaded in the galaxy inner regions than in the outer ones.  It is
known that the X-ray emission from the wind regions is not due to a
volume-filling wind fluid, but rather it arises at the interacting
boundaries of the wind and of the ambient material \citep[][and
references within]{marcolini05}. Thus the metallicity of the X-ray
emitting gas results from the mixing of the wind with the ambient
material.

The wind/cloud interaction is related to the mass loading of the wind,
i.e.\ whether the clouds survive to the wind encounter, or on the
contrary if they are disrupted, their material is accelerated and
becomes part of the wind.  The lifetime of the clouds is not very well
known, since it depends on how thermal conduction is modelled at the
wind/cloud border.  \citet{marcolini05} estimate for a cloud of radius
$R=15$ pc a lifetime in the 0.9--12.8 Myr interval, depending on the
conduction model. It is also known, however, that it scales with
$R^2$, $R$ being the cloud radius, so that larger clouds may have much
longer lifetimes.

Thus, recalling that the starburst age in M82 is $\lesssim 25$ Myr, it
is possible that in the inner regions on M82 there are still some
clouds not yet disrupted.  If the total surface of the clouds which is
exposed to the wind has increased with time, e.g.\ by fragmenting,
than it may be possible that mass loading is becoming more and more
important.  Provided that the supernova-originated wind is enriched
with respect to the ambient gas, and that the wind height on the
galaxy plane may still be related to the elapsed time, the result is
that the metallicity of the X-ray emitting gas is tending towards
lower abundances as one moves from the external to the inner regions.

Two possible objections to this picture may be the contradictory
behaviour of S, and the fact that fragmenting may be not effective in
increasing the clouds surfaces. E.g., if a spherical cloud fragments
in $N$ smaller spherical clouds and the density does not change, than
the total surface $S$ scales with $S\propto N^{1/3}$.

\subsection{Energetics of Charge-Exchange emission}
From a \suzaku\ observation, \citet{tsuru07} calculated that the CE
process can significantly contribute to the OVII $\lambda$21--22 \AA\
lines observed from the region of extended emission known as `the
Cap'. This emission is considered a part of the M82 outflow, and lies
$\sim 11^\prime$ (11.6 kpc) north of the M82 nucleus.  The
best-fitting spectral models of the Cap give a density for the hot gas
of $\sim 5\e{-3}$ cm$^{-3}$. Under the (extreme) hypothesis that all O
ions suffer the CE process and emit O K lines, and assuming a density
of the order $10^{-3}$ cm$^{-3}$ for the neutral material, Tsuru et
al.\ estimated an upper limit to the contribution from CE of $5\e{-6}$
photons s$^{-1}$ cm$^{-2}$, to be compared with an observed flux of
$6\e{-6}$ photons s$^{-1}$ cm$^{-2}$.  To make a similar estimate for
the M82 central region, one has to rescale for the volume and
densities. The volume and the hot gas density are taken from
Table~\ref{tab:fisica} (1.4 kpc$^3$ and $32\e{-3}$ cm$^{-3}$).  The
mass and density of the molecular material contained in the same
volume are $3.7\e{8}$ M$_\odot$ and 11 cm$^{-3}$
\citep{youngscoville84}. The rescaled upper limit on the expected flux
would thus be $4.0\e{-2}$ photons s$^{-1}$ cm$^{-2}$, to be compared
with the observed flux for the OVII triplet which is $2.4\e{-4}$
photons s$^{-1}$ cm$^{-2}$. Thus the flux of the OVII triplet can be
fully justified if the CE process involves $\lesssim 1\%$ of the OVIII
ions. Of course, this is still an upper limit, as the flux contributed
to the triplet by thermal emission is highly uncertain.

\subsection{Reliability of Oxygen abundances. Discrepancies between
  measurements in stars and gas}
\label{sec:disc:ossigeno}
Taken at its face value, the X-ray derived O abundance
is found to be smaller than the ones derived for Fe, Mg and other heavier
$\alpha$-elements.  This effect is true across all the outflow,
although it seems more severe in the central regions: using EPIC
measurements, the O/Mg abundance ratio is $0.7\pm 0.3$ in the external
outflow, and $0.2\pm 0.1$ in the inner part.  On the other hand, WW95
calculates a O/Mg abundance ratio of $\sim 10$--20 for all stars with
masses between 11 and 40 $M_\odot$.  Similarly, the observed O/Si
ranges from 0.2 and 1.0 while WW95 has 2--10.  This was already
noticed by \citet{ume02}, who found in ASCA data O/Mg$\sim 0.2$ and
O/Si$\sim 0.15$. \citet{ume02} had some success in explaining these
abundance ratios by invoking hypernova explosions, i.e.\ energetic
($E\gtrsim 10^{52}$ erg) core-collapse supernovae with massive
($M\gtrsim 25 M_\odot$) metal-poor ($Z=0$) progenitors.

However, the X-ray derived oxygen absolute abundance is smaller than
the IR-derived one. This looks like a more severe problem, because the
IR-derived abundances should trace the composition of the gas before
any enhancement by the last starburst occurs. 

A significant contribution from CE emission has been found in the O
VII lines (Sect.~\ref{sec:righe-O}), thus challenging also the O VIII
lines because CE is expected to be found in lines from both ionisation
states \citep{djuric05}. While this might question the reliability of
O abundance measurements from X-ray observations, the contributions
from CE and thermal emission should add together and result in a
stronger line, and hence in a larger measured abundance. Thus it seems
that the problem of O under-anbundance is only made worse.

\citet{marcolini05} found, in simulation of wind/cloud interaction,
that the conductive interface between wind and clouds has a lower
temperature than the wind bow shock.  Thus they suggested that, since
the emissivity of different ions is strongly temperature dependent,
the X-ray line emission from low mass ions (O VII, O
  VIII) should be more influenced by cloud material, while the
contribution from wind material should be larger for higher ionisation
species (Mg and heavier).  On top of this, one could also consider the
CE as a cooling mechanism for O ions, because in reactions between O
and an element $X$ like
\begin{equation}
  \label{eq:CE}
  O^{q+}+X\rightarrow O^{(q-1)+*}+X^+
\end{equation}
the O ion looses energy by falling in a lower ionisation state, and
may eventually end out of the `temperature window' available for
emission in the 0.5--2.0 keV band. So the bulk of the O mass could
probably be traced by measuring O in the UV and far UV bands.

\section{Conclusions}
\label{sec:conclusioni}
We have reported on a deep ($\sim 100$ ks) X-ray observation of the
starburst galaxy M82 conduced with the \xmm\ satellite. Spectra of
both the galaxy centre and the outflow have been presented. 

By observing the M82 nuclear regions with EPIC, at least three
spectral components were found: a power-law making most of the flux at
energies larger than a few keV, a collisionally ionised plasma
accounting for most of the flux at lower energies, and three narrow
lines whose origin may be attributed to charge exchange emission. 

The power-law emission may be ascribed to the many point sources
present in the M82 central region, among whom M82-X1 is by far the
most luminous. Its data from this observation have been reported
elsewhere \citep{mucciarelli06}.

The thermal emission has been found to have a double-peaked
differential emission measure, the first peak being at $\sim 0.5$ keV
and the second one at $\sim 7$ keV. A similar temperature structure,
albeit with a smaller temperature ratio, is predicted by the model
proposed by \citet{masaidogiel02}. According to these authors, if a
plasma is subject to particle acceleration co-spatial with the thermal
emission, then the energy spectrum of the electrons is not fully
thermal but includes a non-thermal part, simulating a hot component in
the X-ray photon spectrum.  Indeed, the Fe line at $\sim 6.7$ keV is
consistent with a plasma with temperature $kT\sim 0.5$ keV but not
with a temperature of $\sim 7$ keV.

The double-peaked structure of the DEM was confirmed in a series of
spectra of the outflow taken with EPIC at increasing distance from the
galactic plane. The plasma temperatures do not change significantly,
the lower energy peak moving at lower energies further from the galactic
plane, while the upper energy peak stays constant across the centre
and all the outflow.

Three lines were found in the EPIC spectra which may be attributed to
charge exchange reactions involving neutral Mg and Si present in dust
grains and the O VIII $n=3\rightarrow 1$ transition. The equivalent
width of these lines depend on the camera (MOS or {\em pn}) and on the
spectral model used for the plasma (MEKAL or APEC), ranging from $\sim
4$ to $\sim 40$ eV.  The RGS data confirm the CE lines from Mg and Si,
while the CE line from O is not confirmed. RGS data also hint for a CE
origin of the O VII $\lambda 21.6$--22.1 triplet (EPIC cannot place
constraints on the line intensity ratios of this triplet because of
insufficient spectral resolution).

Both the O VIII line at 19 \AA\ and the O VII triplet at 21--22 \AA\
consist of two components: the main one, at a redshift similar to that
of the other elements' lines, and a secondary one with lower intensity
which exhibits a blueshift ($z\sim -0.012$ corresponding to $\Delta
v\sim 3800$ km s$^{-1}$). Moreover, the lines making the dominant
(redshifted) component O VII triplet have line intensity ratios
consistent with CE emission but not with a thermal plasma.

The chemical abundances of O (taken at its face value and ignoring any
possible contribution from CE), Ne, Mg, Si and Fe as derived with EPIC
increase when moving out of the galaxy centre and towards the external
regions of the outflow, while the behaviour of S is unclear. On the
other hand, since the RGS is a slitless spectrometer with a large
field of view, the spectral properties derived with this instrument
should be regarded as field-of-view-averaged. Indeed, the RGS-derived
abundances are consistent with being a spatially-weighted average of
the EPIC ones.  Abundance ratios, such as Mg/Fe or Si/Fe also show
spatial structure.

Confirming our previous study \citep{or04}, the X-ray derived O
abundance in the M82 centre is lower than the IR-derived one, at
variance with the expectation that the X-ray emitting gas would be
enriched with respect to the stars. Reminding that the X-ray emission
from different elements preferably occurs at different temperatures,
and that the main site for X-ray emission is the interface between the
hot gas and the colder ambient material \citep{marcolini05}, we suggest
that a significant fraction of the O VIII and O VII ions already
have cooled off through interaction with cold material or the CE
mechanism. Observations in the far ultraviolet regime could help
clarify this issue.

\section*{Acknowledgments}

The present version of this paper greatly benefited from a detailed
and constructive report from an anonymous referee, who is warmly
thanked for her/his job.

This work would not have been possible without the kindness and
support of Prof.\ K. Makishima, director of the Cosmic Radiation
Laboratory of the RIKEN institute, who allowed P.R. to work
continuously on this project.  P.R. would also like to thank A. Bamba,
P. Gandhi, N. Isobe, N. Ota, A. Senda, M. Suzuki, Y. Terada, and all
the staff of the above laboratory. We thank T. Tsuru and
E. Costantini for valuable help and constructive discussions, and
K. Arnaud for support of the XSPEC package.

This work was supported by a fellowship from the Japan Society for the
Promotion of Science, and by two grants from the RIKEN institute.
Part of this work was also supported by ASI/INAF contract I/023/05/0
and by PRIM/MUR grant 2006-02-5203.

%
%
%

\def\aj{AJ}%
\def\araa{ARA\&A}%
\def\apj{ApJ}%
\def\apjl{ApJ}%
\def\apjs{ApJS}%
\def\ao{Appl.~Opt.}%
\def\apss{Ap\&SS}%
\def\aap{A\&A}%
\def\aapr{A\&A~Rev.}%
\def\aaps{A\&AS}%
\def\azh{AZh}%
\def\baas{BAAS}%
\def\jrasc{JRASC}%
\def\memras{MmRAS}%
\def\mnras{MNRAS}%
\def\pra{Phys.~Rev.~A}%
\def\prb{Phys.~Rev.~B}%
\def\prc{Phys.~Rev.~C}%
\def\prd{Phys.~Rev.~D}%
\def\pre{Phys.~Rev.~E}%
\def\prl{Phys.~Rev.~Lett.}%
\def\pasp{PASP}%
\def\pasj{PASJ}%
\def\qjras{QJRAS}%
\def\skytel{S\&T}%
\def\solphys{Sol.~Phys.}%
\def\sovast{Soviet~Ast.}%
\def\ssr{Space~Sci.~Rev.}%
\def\zap{ZAp}%
\def\nat{Nature}%
\def\iaucirc{IAU~Circ.}%
\def\aplett{Astrophys.~Lett.}%
\def\apspr{Astrophys.~Space~Phys.~Res.}%
\def\bain{Bull.~Astron.~Inst.~Netherlands}%
\def\fcp{Fund.~Cosmic~Phys.}%
\def\gca{Geochim.~Cosmochim.~Acta}%
\def\grl{Geophys.~Res.~Lett.}%
\def\jcp{J.~Chem.~Phys.}%
\def\jgr{J.~Geophys.~Res.}%
\def\jqsrt{J.~Quant.~Spec.~Radiat.~Transf.}%
\def\memsai{Mem.~Soc.~Astron.~Italiana}%
\def\nphysa{Nucl.~Phys.~A}%
\def\physrep{Phys.~Rep.}%
\def\physscr{Phys.~Scr}%
\def\planss{Planet.~Space~Sci.}%
\def\procspie{Proc.~SPIE}%
\let\astap=\aap
\let\apjlett=\apjl
\let\apjsupp=\apjs
\let\applopt=\ao
\bibliographystyle{mn}
\bibliography{fullbiblio}

\label{lastpage}
\end{document}